\documentclass[aps,prl,twocolumn,superscriptaddress,groupedaddress]{revtex4-1}
\usepackage[compat=1.0.0]{tikz-feynman}

\usepackage{amssymb}
\usepackage{multirow}
\usepackage{bm}
\usepackage{amsmath}
\usepackage{graphicx}
\usepackage{epstopdf}
\usepackage{subfigure}
\usepackage{natbib}
\usepackage{epsfig}
\usepackage{amsfonts}
\usepackage{mathrsfs}
\usepackage[toc,page,title,titletoc,header]{appendix}
\usepackage[colorlinks,linkcolor=blue,citecolor=blue,anchorcolor=blue]{hyperref}

\usepackage{dsfont,amsthm,amsbsy}

\usepackage{fancyhdr}
\usepackage{ulem}

\newcommand{\bea}{\begin{equation} \begin{aligned}}
\newcommand{\eea}{\end{aligned} \end{equation} }

\usepackage{braket}

\def\Ueph{U_\text{e-ph}}

\begin{document}

\title{``Quantum Bipolaron'' Superconductivity from Quadratic Electron-Phonon Coupling}
\author{Zhaoyu Han}
\affiliation{Department of Physics, Stanford University, Stanford, California 94305, USA}

\author{Steven A. Kivelson}
\affiliation{Department of Physics, Stanford University, Stanford, California 94305, USA}

\author{Pavel A. Volkov}

\affiliation{Department of Physics, University of Connecticut, Storrs, Connecticut 06269, USA}
\affiliation{Department of Physics, Harvard University, Cambridge, Massachusetts 02138, USA}

\begin{abstract}
When the electron-phonon coupling is quadratic in the phonon coordinates,  electrons can pair to form bipolarons due to phonon zero-point fluctuations, a purely quantum effect. We study superconductivity originating from this pairing mechanism in a minimal model and reveal that, in the strong coupling regime, the critical temperature ($T_c$) is only mildly suppressed by the coupling strength, in stark contrast to the exponential suppression in linearly coupled systems, thus implying higher optimal $T_c$ values. We demonstrate that large coupling constants of this flavor are achieved in known materials such as perovskites, and discuss strategies to realize such superconductivity using superlattices.
\end{abstract}

\maketitle

The electron-phonon (e-ph) interaction plays an essential role in many quantum materials that exhibit superconductivity (SC)~\cite{PhysRev.108.1175,Marsiglio2008,Carlson2004}. It is generally assumed that pairing primarily arises from linear couplings between electron densities and phonon coordinates. In this conventional setup, it has long been recognized that the superconducting critical temperature ($T_c$) is small both for large and small values of the dimensionless electron-phonon coupling, $\lambda\equiv\rho_0 \Ueph$, where $\Ueph$ is the characteristic energy scale of phonon-induced attraction between two electrons, and $\rho_0$ is the density of states at the Fermi energy, $E_F$. In the weak coupling, Bardeen-Cooper-Schrieffer (BCS) limit, this reflects an exponentially small pairing scale, $\Delta \sim \exp[-1/\lambda]$, while for strong coupling regime, $T_c$ is set by the condensation temperature of Cooper pairs (preformed bipolarons), which is inversely proportional to their parametrically heavy effective mass $m^\star \sim \exp[2\Ueph/(\hbar\omega_0)]$, where $\omega_0$ is a characteristic phonon frequency~\cite{Carlson2004,ChakravertyBipolaron, chakraverty1987squeezed, PhysRevLett.84.3153,alexandrov2001breakdown,esterlis2018breakdown,esterlis2019pseudogap,PhysRevLett.125.167001,PhysRevB.48.3881}. The maximum $T_c$ for the (often realistic) case $\hbar \omega_0\ll E_F$ has been estimated (on the basis of numerics) to arise for $\lambda \sim 1$, where $k_BT_c$ is a small fraction ($\sim 0.1$) of $\hbar \omega_0$~\cite{ChakravertyBipolaron, CohenBounds,esterlis2018bound,hofmann2022boundTc,Chubukov}. (This heuristic bound could be violated in models with a large number of comparably strongly coupled phonon modes~\cite{hofmann2022heuristic}, or when the phonon couples to the electron hopping matrix elements~\cite{zhang2023bipolaron,SousCoulomb, wang2022robust,tanjaroon2023comparative,CarboneBond,BoSSH2D,PhysRevLett.127.247203,ScaletterSSHU,PhysRevB.106.L081115,PhysRevB.105.085151,PhysRevLett.130.186404,gotz2023phases,costa2023comparative,kim2023semi})

In this Letter, we consider e-ph couplings that are quadratic in the phonon coordinates and linear in the electron density introduced previously in context of various systems~\cite{PhysRevLett.32.215,kuklov1989soliton,ENTINWOHLMAN19831,RISEBOROUGH19841,Entin-Wohlman_1985, HIZHNYAKOV2010191, PhysRevB.45.5052,PhysRev.168.936,PhysRevB.92.064301,gogolin1991quantum,PhysRevB.89.035122,PhysRevB.90.085149,PhysRevB.56.8322,adolphs2013going,kiseliov,volkov2022superconductivity,PhysRevLett.87.037001,PhysRevB.104.L220506,PhysRevResearch.1.013003,kumar2021,PhysRevB.107.L121109,zhang2023soliton} but for which, to date, the strong-coupling regime and optimal $T_c$'s have not been considered. We find that this type of coupling leads to the formation of small bipolarons by a purely quantum mechanical effect - a reduction of the zero-point energy of the phonons, without any accompanying lattice displacement. As a result, the exponential mass enhancement characteristic of the linear problem is replaced by a much weaker, polynomial mass enhancement, $m^\star \sim \Ueph^{3/2}$. Moreover, even in the extreme strong-coupling limit ($\Ueph\to \infty$), where charge density wave (CDW) order always precludes SC in the linearly coupled case, in the present case we find a finite range of densities in which the ground state is SC. These results suggest higher optimal $T_c$ values than achievable with linear couplings. We theoretically estimate the strengths of quadratic e-ph coupling in real materials and show that large coupling strengths saturating the estimate are realizable in real materials. We also show that engineered two-dimensional (2D) superlattices can help to achieve strong coupling SC of this kind and potentially lead to high $T_c$ values.

{\bf The Model.} Studies of the Holstein model~\cite{holstein1959studies} have led to significant advances in the understanding of the generic physics of the electron-phonon system in real materials, despite its simple form~\cite{PhysRevB.65.174306,PhysRevB.60.1633,aubry1992chaotic,alexandrov2010advances,trugman1999polaron,PhysRevB.61.12315,PhysRevB.75.014307,PhysRevB.100.094307,PhysRevB.58.3094,barivsic2006quantum,PhysRevLett.82.807,PhysRevB.106.155129,fehske2007numerical, PhysRevB.107.075142,barivsic2008phase,PhysRevLett.109.236402}. Following the same spirit, in this work, we study a direct generalization, the quadratic Holstein model~\cite{PhysRevB.107.L121109,zhang2023soliton}:
\begin{align}
    \hat{H} = & -t \sum_{\langle ij \rangle \sigma } \left(\hat{c}^\dagger_{i\sigma}\hat{c}_{j\sigma} + \text{h.c.}\right) \nonumber\\
    & \ \ + \sum_{i} \left[\frac{K}{2}\left( \hat{X}_i^2+g \hat{n}_i\hat{X}_i^2\right) +\frac{\hat{P}_i^2}{2M}\right]
\end{align}
where $\hat{c}_{i\sigma}$ annihilates a spin-$\sigma$ electron on site-$i$, $\hat{n}_i=\sum_\sigma \hat{c}^\dagger_{i\sigma}\hat{c}_{i\sigma}$ is the electron density, $\hat{X}_i$ and $\hat{P}_i$ are the coordinate and momentum operators of the optical phonon, $K$ and $M$ are the bare stiffness and ion mass, and $g$ is a dimensionless coupling constant. It must be assumed that $g>-1/2$ for the stability of this model. (When $g<-1/2$, higher order terms in the phonon potential must be included.) On a site with $m=0,1,2$ electrons, the phonon oscillates with frequency $\omega_m \equiv \omega_0\sqrt{1+mg} $ where $\omega_0 \equiv \sqrt{K/M}$ is the bare phonon frequency of the system. Below we will show that $\Ueph=\hbar\omega_1 - \hbar(\omega_2+\omega_0)/2$.

The more familiar (linear) Holstein model, to which we will make comparisons, is of the same form but with $Kg\hat{n}_i \hat{X}_i^2 \to -2\alpha \hat{n}_i \hat{X}_i$, and $\Ueph=\alpha^2/K$. 

\begin{figure}[b!]
    \centering
    \includegraphics[width=\linewidth]{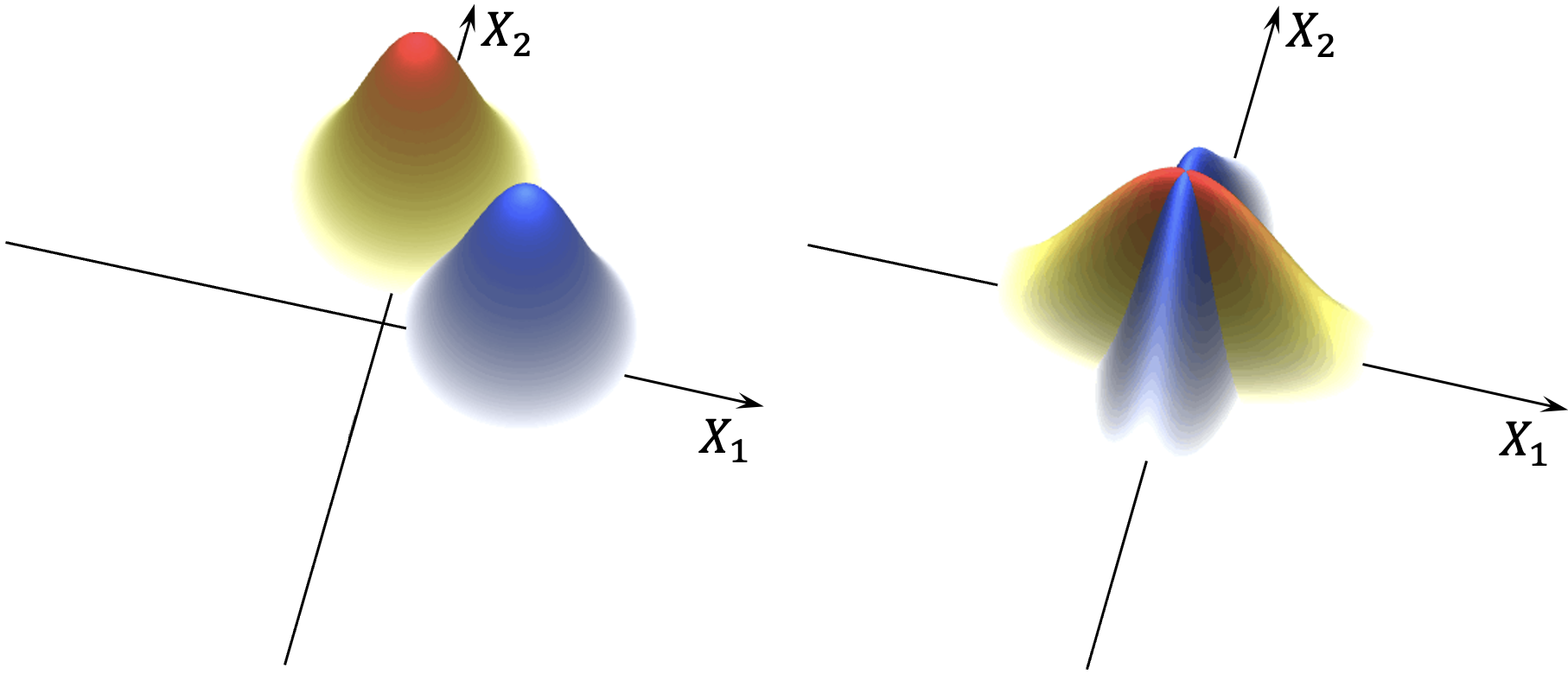}
    \caption{Illustration of the phonon wavefunction amplitude before (in blue) and after (in red) a bipolaron hops from site $1$ to site $2$, whose phonon coordinates are $X_1$ and $X_2$. The left panel shows the conventional case of linear e-ph coupling, where the phonon equilibrium positions are displaced during the process; the right panel shows the quadratic case, where only the spreads of the wavefunctions change. The overlaps between the wavefunctions essentially result in the suppression of the effective hopping amplitude of bipolaron, which is exponentially small in $\Ueph$ in the linear case but is only polynomially small in the quadratic case.
    }
    \label{fig: illustration}
\end{figure}

{\bf ``Quantum bipolarons.''}  To understand the origin of the effective electron-electron attraction, consider the atomic limit where $t=0$. Since now the number of electrons on each site is conserved, we can evaluate the effective interaction between a pair of electrons by comparing the ground-state energy when they are placed on two distinct sites, or both placed on the same site. As illustrated in Fig.~\ref{fig: illustration}, in the linear Holstein model, the equilibrium value of the phonon coordinate depends on the occupancy of the site, $\langle \hat{X}_i\rangle =  \langle \hat{n}_i \rangle \alpha/K$, and correspondingly there is an effective bipolaron binding energy $\Ueph=\alpha^2/K$ that is classical in the sense that it is independent of $M$, even as $M\to\infty$. For the quadratic Holstein model, $\langle \hat{X}_i\rangle=0$ is independent of the electron occupation number;  however, the phonon quantum zero-point energy is occupation-number-dependent as long as $M$ is finite. Specifically, the energy of one doubly occupied site and one empty site is lower than that of two singly occupied sites by an amount
\begin{align}
   \Ueph
    = & \hbar\omega_0\left(\sqrt{1+g}-\frac{\sqrt{1+2g}+1}{2}\right)
\end{align}  

Importantly, the energy gain of binding two electrons together is always positive for any $g>-1/2$ (since $\omega_1 = \sqrt{(\omega_0^2+\omega_2^2)/2}>(\omega_0+\omega_2)/2$). The origin of this attraction is purely a quantum mechanical effect that is intrinsically different from that of the linear e-ph coupling; for this reason, we call the bipolarons formed by this mechanism ``quantum bipolarons'' \footnote{A similar theory at {\it finite} phonon densities has been proposed to explain certain light-induced transient pairings~\cite{kennes2017transient,sous2021phonon}.}. 

{\bf Weak coupling limit: } When $g\rightarrow 0$, the characteristic energy scale, $\Ueph\sim \hbar\omega_0 g^2/8$, 
appears as the effective interaction vertex in the diagrammatic treatment~\cite{kiseliov,volkov2022superconductivity}. As long as $\lambda = \rho_0\Ueph \ll 1$, the standard BCS analysis applies, and we obtain the familiar expression for $T_c$ 
\cite{gorkov2016,chubukov2016}: 
\begin{align}\label{eq: weak coupling Tc}
    T^{\text{BCS}}_c \sim \min(\hbar \omega_0,E_F)  \mathrm{e}^{- 1/(\rho_0 \Ueph)},
\end{align}
where $E_F$ is the Fermi energy and $\rho_0$ is the density of states at the Fermi level. One interesting case is small electron density, $n\ll 1$, where $E_F\sim n^{2/d} |t| $, $\rho_0 \sim n^{1-2/d} / |t|$ and where $d$ is the spatial dimension. Despite its formal similarity to the results in the usual Holstein model, we note that this formula implies an anomalously {\it strong} isotope effect since $\Ueph\sim 1/\sqrt{M}$.

\begin{figure}[t!]
    \centering
    \includegraphics[width=\linewidth]{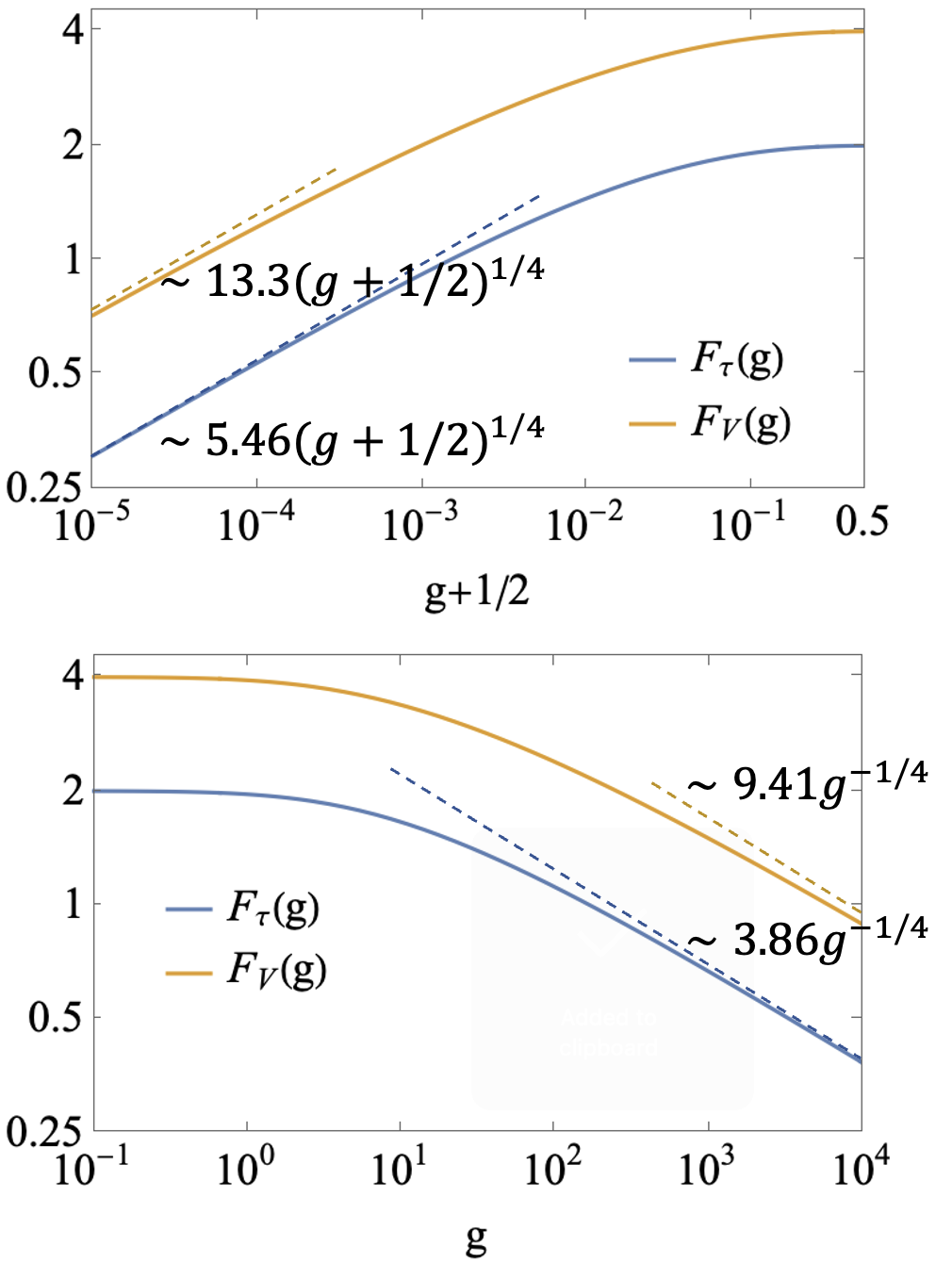}
    \caption{The log-log plot of $F_\tau(g)$ and $F_V(g)$ (as defined in Eqs.~\ref{eq: Ftau}\&\ref{eq: FV}) for $g+1/2 \in [0,1/2]$ (upper panel) and $g\in [0,\infty]$ (lower panel). The asymptotic scaling behaviors as $g$ approaches the two extremes are indicated by dashed lines. }
    \label{fig: F}
\end{figure}

{\bf Strong coupling limit:} We next analyze the problem in the ``strong-coupling'' limit, $ \Ueph \gg |t|$. To the zeroth order in $t$, the degenerate ground space manifold consists of different occupation configurations of quantum bipolarons (with no phonons). Within this subspace, we then perform a perturbative expansion in powers of $t$ to obtain a low-energy effective model. To the second order, the resulting Hamiltonian has the same form as that for the conventional Holstein model, i.e. it maps to a model of {\it hard-core} bosons (bipolarons), with annihilation operators $\hat{b}_i \equiv \hat{c}_{i\uparrow} \hat{c}_{i\downarrow}$:
\begin{align}\label{eq: effective model}
    \hat{H}_\text{eff} = -\tau \sum_{\langle ij\rangle} \left(\hat{b}^\dagger_i\hat{b}_j + \text{h.c.}\right) + V \sum_{\langle ij\rangle} \hat{b}^\dagger_i\hat{b}_i\hat{b}^\dagger_j\hat{b}_j
\end{align}
However, the expressions for $\tau$ and $V$, derived (explicitly in the Supplemental Material \footnote{See Supplemental Material [url] for the derivation of effective coefficients and the estimate of $g$ in for STO, and Refs.~\cite{vanderbilt2001,wannierrev,bauerle1980soft} therein.}) by summing over virtual processes associated with intermediate states with all possible phonon excitations, are crucially different than the corresponding expressions for the linear Holstein model. The results can be expressed as 
\begin{align}
 \tau =\frac{t^2}{\Ueph} F_\tau(g) \ ; \ \ \ V =\frac{t^2}{\Ueph} F_V(g)\ ,
\end{align}
\begin{align}
    F_\tau(g) &= 2\eta \int_0^\infty dz \frac{ \mathrm{e}^{-z}}{1-\gamma_0\gamma_2\mathrm{e}^{-2\xi z}}\label{eq: Ftau} \\
    F_V(g) &= 4\eta \int_0^\infty dz \frac{ \mathrm{e}^{-z}}{\sqrt{1-\gamma_0^2\mathrm{e}^{-2\xi z}}\sqrt{1-\gamma_2^2\mathrm{e}^{-2\xi z}}} \label{eq: FV}
\end{align}
where
\begin{align}
    \xi &\equiv \frac{2\sqrt{1+g}}{2\sqrt{1+g}-\sqrt{1+2g}-1} \label{eq: xi}\\
    \gamma_{m} & \equiv \frac{\sqrt{1+g}-\sqrt{1+mg}}{\sqrt{1+g}+\sqrt{1+mg}} \\
    \eta &\equiv \frac{4\sqrt[4]{1+2g} \sqrt{1+g} }{\left(1+\sqrt{1+g}\right)\left(\sqrt{1+2 g}+\sqrt{1+g}\right)}
\end{align}
These expressions are the central results of this work. Their dependence on $g$ is plotted, and their asymptotic behaviors in the large $g$ and $g\rightarrow -1/2$ limits are indicated in Fig.~\ref{fig: F}.

The most important feature is that $F_\tau(g)$ is only polynomially suppressed in the strong coupling limit, in stark contrast to the cases with linear e-ph couplings for which the suppression of $\tau\sim 1/m^\star$ is exponentially strong $\sim \exp[-2\Ueph/(\hbar\omega_0)]$ ~\cite{Carlson2004,PhysRevLett.125.167001,PhysRevB.48.3881}. This can be easily understood by recognizing that when a bipolaron hops from one site to another, no phonon needs to be classically displaced (see Fig. \ref{fig: illustration}). Therefore, the overlap between initial and final phonon wavefunctions is substantial, in contrast to the linear case.

The hard-core boson model in Eq.~\ref{eq: effective model} (or equivalently the spin-$1/2$ XXZ model) has been extensively studied on various lattices and dimensions~\cite{PhysRevB.56.3212,PhysRevLett.84.1599,PhysRevLett.88.167208,PhysRevB.84.054510,PhysRevB.77.014524,PhysRevLett.95.127207,PhysRevLett.102.017203,PhysRevB.75.174301,PhysRevLett.112.127203,PhysRevLett.97.147202,PhysRevLett.95.237204,PhysRevB.91.081104,PhysRevB.84.174515,PhysRevA.89.013615,spevak2021critical}. The nature of the low $T$ phases generically depends on $\tau/V$ and the boson density $n_\text{bipolaron} =n/2$. At dilute densities $n_\text{bipolaron} < n_c$ and at low temperatures $T<T_c \sim  n_\text{bipolaron}^{2/d}\ \tau$, SC generically develops~\cite{PhysRevB.37.4936, PhysRevLett.87.270402,PhysRevLett.100.140405} (even in the presence of an additional long-ranged Coulomb repulsion as long as the density is not extremely low~\cite{PhysRevLett.130.236001}). The critical density, $n_c$ (to the formation of some form of commensurate CDW order with phase separation) depends on the lattice structure but generally is an increasing function of $\tau/V$. Generically, as long as $\tau/V$ is not too small, SC can be stable in a broad density range (even for all densities on several frustrated lattices~\cite{PhysRevB.84.174515,PhysRevLett.97.147202}). In the linear Holstein model, $\tau/V \rightarrow 0$ rapidly with increasing coupling. In the quadratic case, on the contrary,  $\tau/V$ never approaches zero, even when $g\rightarrow \infty$. More quantitatively, $\tau/V$ varies from $0.5$ to $0.4$ as $g$ varies from $0$ to $\infty$ or $-1/2$. Given this lower bound on $\tau/V$, $n_c$ remains finite for the whole strong coupling regime (for example, $n_c \gtrsim 0.4$ for square lattice~\cite{spevak2021critical} and $n_c = 1$ for triangular lattice~\cite{PhysRevB.84.174515}).

For dilute $n=2n_\text{bipolaron}<2n_c $, the SC transition temperature can be estimated as
\begin{align}\label{eq: strong coupling Tc}
    T_c \sim n^{2/d} \frac{t^2}{\Ueph} g^{-1/4}.
\end{align}
This implies a remarkable ``inverse isotope effect'' at strong coupling, reversing the trend at weak coupling: $T_c$ is proportional to the square root of ion mass, $T_c\propto \sqrt{M}$ (holding all the other parameters fixed)!

We note that unusually weak polaron mass enhancement has been numerically observed in the {\it single-electron} (polaron) sector of the same model in Refs.~\cite{PhysRevB.107.L121109,zhang2023soliton}. It is straightforward to show with a similar strong-coupling analysis that the mass enhancement in this case $\sim g^{3/4}$. Our results imply that at finite electron densities, the polaron liquid is unstable to bipolaron formation, leading to an ordered many-body ground state. We also note that mass enhancement of polarons and bipolarons for {\it negative} $g$ has been explored in Refs.~\cite{PhysRevB.89.035122,PhysRevB.90.085149} where large $|g|$ (regulated by a quartic term in the phonon potential energy) have been shown to lead to exponential mass suppression as in the usual, linear case.

We remark that the perturbation series presented in this work can be calculated to higher orders in $t/\Ueph$ in a systematic manner. The subleading terms include further ranged effective bipolaron hoppings and interactions, reflecting corrections to the binding energy and the effective bipolaron radius.

\begin{figure}[t!]
    \centering
    \includegraphics[width= \linewidth]{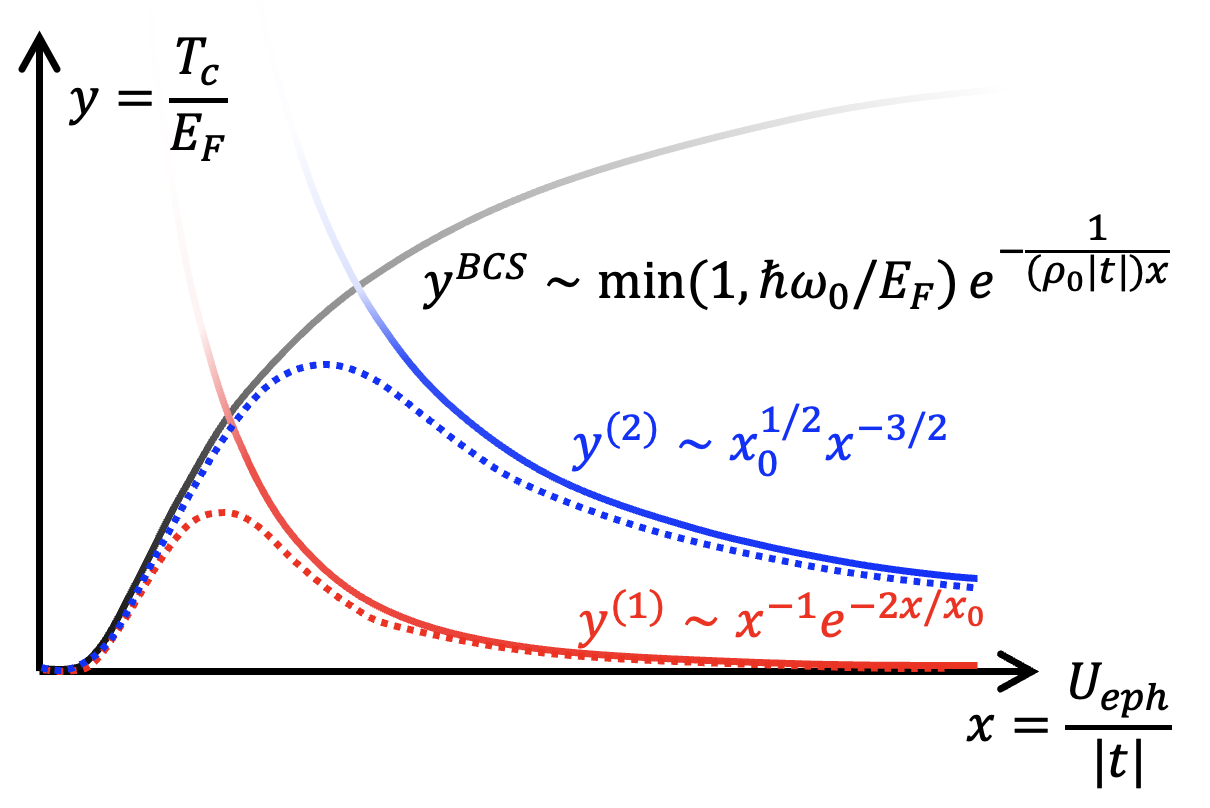}
    \caption{Schematic representation of the dimensionless measures of the mean-field  (BCS) collective pairing scale  ($ y^{(BCS)}$ - the solid black line) and the strong coupling condensation scale for the quadratic  ($y^{(2)}$ - the blue solid line) and linear e-ph systems ($y^{(1)}$ - the red solid line) with dilute ($n \ll 1$) electrons.  $y=T/E_F$ is the dimensionless temperature and $x=\Ueph/|t|$ is the dimensionless coupling strength. The parameter $x_0 = \hbar\omega_0/ |t|$ also enters these expressions. The dashed curves show a plausible interpolation between the limiting behaviors for the two models. The asymptotic expressions for $y^{(1)}$ and $y^{(2)}$ are from, respectively, Refs.~\cite{Carlson2004,PhysRevLett.125.167001,PhysRevB.48.3881} and Eq.~\ref{eq: strong coupling Tc}. }
    \label{fig: comparison}
\end{figure}

{\bf Discussion: } In both the weak and strong coupling limits, we have obtained well-controlled estimates of the SC $T_c$, corresponding to the pairing and phase coherence scales in the two limits, respectively.  In the intermediate coupling regime, both factors together determine the physical $T_c$, the maximum value of which could thus be reached by tuning the interaction strength to a ``sweet spot'' interpolating the two asymptotic behaviors. Since the weak coupling side is described by BCS theory for both linear and quadratic e-ph couplings, it is crucial to understand the enhancement of optimal $T_c$ from the strong coupling side. To illustrate the difference, in Fig.~\ref{fig: comparison} we plot the schematic behavior of $T_c$ for both quadratic and linear Holstein models. Because SC is so much stronger in the strong coupling limit in the quadratic coupling case, it is certainly plausible (as represented by the dashed lines interpolating between the controllable limits in the figure) that the optimal $T_c$ is substantially higher.

Importantly, our central results remain robust when relatively weaker linear couplings coexist with quadratic ones, as long as the average phonon displacement associated with bipolaron hoping, $\langle \hat{X} \rangle \sim \frac{\alpha}{(1+2 g) K}$, is small compared to the root mean squared coordinate fluctuations, i.e. $\frac{\alpha}{(1+2 g) K} \lesssim \sqrt{\frac{\hbar}{M\omega_0}}$. Thus, besides a large $g$ and a small $\alpha$, a large phonon stiffness $K$ and a small ion mass $M$ are also conducive to the quadratic e-ph couplings playing a central role.

Turning to the real-world implications, the local quadratic e-ph couplings are ubiquitous in materials, since they are always allowed by symmetry. In contrast, linear on-site coupling to the electron density is forbidden by symmetry for various phonon modes.  An interesting example is a transverse polar phonon. The conventional e-ph gradient coupling vanishes exactly for these modes, as they generate no bound charges. More generally, the fact that the polar phonon is odd under inversion requires any linear coupling to be either interband or intersite~\cite{kozii2015,volkov2020} and consequently suggests that nonlinear couplings may be dominant, especially in the single-band case. Furthermore, in 2D systems, the mirror reflection symmetry along $z$ axis precludes linear e-ph coupling to certain phonons. In particular, out-of-plane optical phonon modes (known as ZO modes) that are odd under the reflection cannot linearly couple to the electron density operators. Such structures are experimentally realizable (e.g. in a magic-angle twisted trilayer graphene~\cite{park2021tunable}), and both $K$ and $gK$ of all ZO modes can be tuned by pressure. 

An estimate of the scale of $\Ueph$ from quadratic e-ph couplings can be obtained as follows: The coupling $gK$ originates from intra-unit cell Coulomb force and therefore the natural energy unit for it is  $E_0\sim\frac{Ze^2}{a_0}$, where $a_0$ is the lattice constant and $Z$ is the phonon Born effective charge.
This leads to $gK\sim E_0/a_0^2$ and thus $\Ueph\sim\hbar\sqrt{gK/M} \sim \sqrt{E_0 \frac{\hbar^2}{M a_0^2}}$. Taking $E_0$ in the range $0.1-1$ Ry for a lattice constant of a few angstroms, and with the ionic mass being $10^{3-4}$ larger than the electron mass, we estimate $\Ueph$ to be as large as order $100$ meV. In fact, in a perovskite, SrTiO$_3$, the value of $gK$ can be estimated from the density-dependent shift of the soft TO phonon frequency~\cite{bauerle1980soft,volkov2022superconductivity}, which implies a large $g\approx 87$ and $\Ueph\sim 50$ meV~\cite{Note2}.

\begin{figure}
   \centering
   \includegraphics[width = \linewidth]{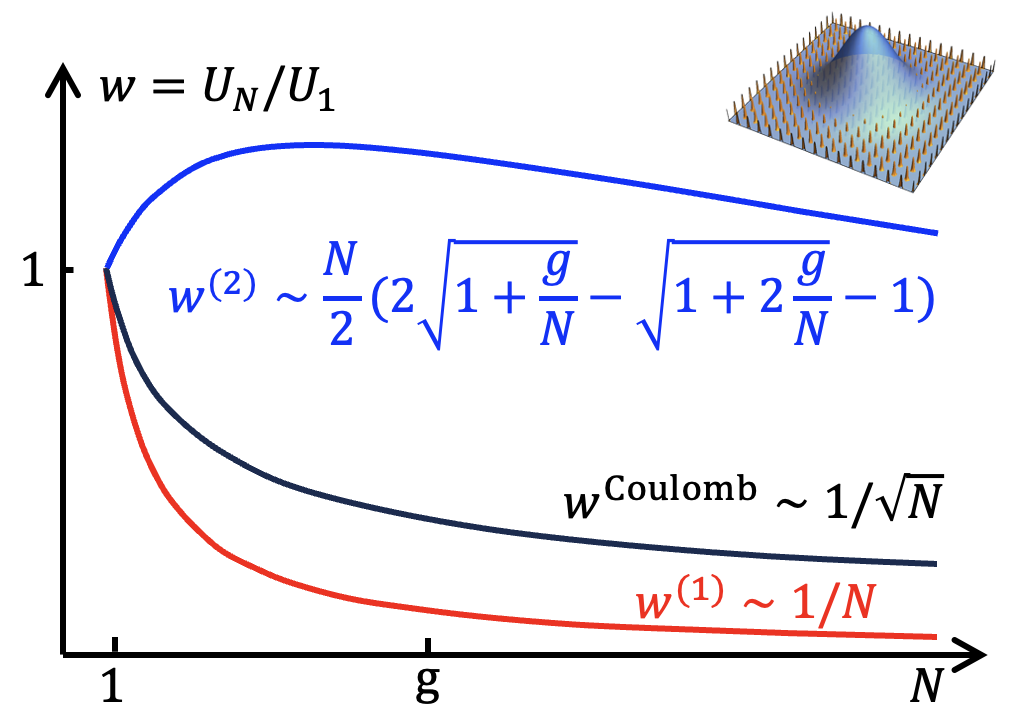}
   \caption{Schematic behavior of the effective interaction strengths in a moir\'e superlattice as a function of $N$, the averaged number of microscopic unit cells overlapping with each Wannier orbital. The strengths are normalized by the case of $N=1$. The curves respectively correspond to direct Coulomb repulsion ($w^\text{Coulomb}$ - the black line), phonon-mediated attraction originating from linear ($w^{(1)}$ - the red line) and quadratic ($w^{(2)}$ - the blue line) e-ph couplings. }
   \label{fig: moire}
\end{figure}

Given that quadratic e-ph systems do not suffer from an exponential depression of the condensation scale, we hope this work points towards a new route to high-temperature SC. However, for any physical proposal to be relevant, three criteria need to be satisfied: 1) the linear couplings must be relatively small as analyzed above; 2) the bandwidth must be comparable to or smaller than $\Ueph$; 3) direct electron-electron Coulomb repulsion (which we have neglected in all the above analysis) must be weaker than $\Ueph$~\footnote{We note that superconductivity originating from certain extended forms of bipolarons may still be stabilized when the Coulomb repulsion is stronger than $\Ueph$~\cite{PhysRevLett.84.3153,proville1999small,proville2000small,PROVILLE1998307,barivsic2012bipolarons,PhysRevLett.125.167001,PhysRevB.64.094507,PhysRevB.80.054301,PhysRevB.85.104520,hague2007superlight,PhysRevLett.98.037002}.}. As discussed above, symmetries can forbid linear coupling to certain phonons, achieving 1). We now show that 2) and 3) can be achieved in 2D systems with superlattice engineering. The presence of a periodic superlattice (created by a moir\'e pattern or electrostatically~\cite{forsythe2018band}) creates bands with reduced bandwidth and enlarged size ($a_\text{orb}$) for the single-electron Wannier orbitals  (Fig.~\ref{fig: moire}). This suppresses the strength of the Coulomb repulsion as $U_\text{Coulomb} \sim E_0 \frac{a_0}{a_\text{orb}}\sim E_0N^{-1/2}$ where $N \sim (a_\text{orb}/{a_0})^2$ is the number of microscopic unit cells over which the Wannier orbital is spread.   On the other hand, the enlarged unit cell includes more ($\sim N$) optical phonons; although each of them couples more weakly ($\sim 1/N$) to an electron in a given Wannier state, the combined effect is an effective attraction,  $\Ueph \approx N\hbar\omega_0 (2\sqrt{1+g/N} - \sqrt{1+2g/N} -1 )/2$, which is slightly enhanced for a range of $N\lesssim g$. (Meanwhile, the values of the dimensionless factors $F_\tau$ and $F_V$ can be relatively weakly $N$ dependent.)~\cite{Note2} By contrast, in the linearly coupled case, the effective phonon-mediated attraction can be similarly estimated as $\Ueph\sim \alpha^2/(NK)$, which is always strongly suppressed by the superlattice. Thus, for 2D materials or interfaces with sufficiently large $g$, appropriately strong superlattice potentials can achieve points 2), 3), and partly 1) simultaneously. Moreover, for large orbital sizes, the electric fields extend far enough out of the plane that screening in a substrate with a high dielectric constant (such as SrTiO$_3$) can significantly reduce the Coulomb repulsion between paired electrons~\cite{liu2021tuning}. Indeed, strong-coupling superconductivity has been suggested to occur in SrTiO$_3$-based 2D nanostructures with strongly suppressed kinetic energy~\cite{cheng2015electron}; our work suggests a path to potentially achieve higher $T_c$ in these systems. Alternatively, bringing surfaces of bulk materials into contact with a twist \cite{inbar2023quantum} can create moir\'e patterns for electrons, doped \cite{10.1063/1.3524198} or residing in an epitaxially grown top layer (such as FeSe~\cite{paul2023}).

In conclusion, we have demonstrated that quadratic e-ph interactions lead to the formation of small, light, ``quantum bipolarons'' in the strong coupling regime. We suggest that this implies higher optimal SC transition temperatures for this mechanism. That relatively strong couplings of this sort are physical is illustrated by the value inferred for SrTiO$_3$ based on experimental data. Finally, we have argued that tunable 2D electronic superlattices provide an excellent platform to reach the optimal and strong coupling regimes while suppressing Coulomb repulsion effects, opening the way for the realization of a new class of strong-coupling superconductors.

{\bf Note Added. } After the publication of this work, we become aware that the continuum limit of the problem, where bipolaron size is much larger than the lattice constant, has been explored by a variational approach in Ref.~\cite{Kuklov1990}. Some qualitative features observed in that regime, such as light bipolaron masses, agree with the ones reported here for lattice-scale (``small'') bipolarons. 

We found that the previous version contains an error in the definition Eq.~\ref{eq: xi} and some subsequent discussions, which are corrected in the current version. We thank the authors of Ref.~\cite{zhang2024superconductivity} for pointing out this mistake.

{\bf Acknowledgement. } We acknowledge helpful discussions with Bai Yang Wang, Srinivas Raghu, Ben Feldman, Jiarui Li, John Sous, Chaitanya Murthy, Piers Coleman, and Premala Chandra. ZH is grateful for insightful discussions during the Polaron Meeting at the Center for Computational Quantum Physics of the Flatiron Institute. ZH was funded, in part, by a QuantEmX grant from ICAM and the Gordon and Betty Moore Foundation through Grant GBMF9616.  ZH and SAK were funded in part by the Department of Energy, Office of Basic Energy Sciences, under Contract No. DE-AC02-76SF00515 at Stanford. PAV acknowledges support from the University of Connecticut OVPR Quantum CT seed grant.

\bibliographystyle{apsrev4-1} 
\bibliography{ref}

\begin{thebibliography}{118}%
\makeatletter
\providecommand \@ifxundefined [1]{%
 \@ifx{#1\undefined}
}%
\providecommand \@ifnum [1]{%
 \ifnum #1\expandafter \@firstoftwo
 \else \expandafter \@secondoftwo
 \fi
}%
\providecommand \@ifx [1]{%
 \ifx #1\expandafter \@firstoftwo
 \else \expandafter \@secondoftwo
 \fi
}%
\providecommand \natexlab [1]{#1}%
\providecommand \enquote  [1]{``#1''}%
\providecommand \bibnamefont  [1]{#1}%
\providecommand \bibfnamefont [1]{#1}%
\providecommand \citenamefont [1]{#1}%
\providecommand \href@noop [0]{\@secondoftwo}%
\providecommand \href [0]{\begingroup \@sanitize@url \@href}%
\providecommand \@href[1]{\@@startlink{#1}\@@href}%
\providecommand \@@href[1]{\endgroup#1\@@endlink}%
\providecommand \@sanitize@url [0]{\catcode `\\12\catcode `\$12\catcode
  `\&12\catcode `\#12\catcode `\^12\catcode `\_12\catcode `\%12\relax}%
\providecommand \@@startlink[1]{}%
\providecommand \@@endlink[0]{}%
\providecommand \url  [0]{\begingroup\@sanitize@url \@url }%
\providecommand \@url [1]{\endgroup\@href {#1}{\urlprefix }}%
\providecommand \urlprefix  [0]{URL }%
\providecommand \Eprint [0]{\href }%
\providecommand \doibase [0]{http://dx.doi.org/}%
\providecommand \selectlanguage [0]{\@gobble}%
\providecommand \bibinfo  [0]{\@secondoftwo}%
\providecommand \bibfield  [0]{\@secondoftwo}%
\providecommand \translation [1]{[#1]}%
\providecommand \BibitemOpen [0]{}%
\providecommand \bibitemStop [0]{}%
\providecommand \bibitemNoStop [0]{.\EOS\space}%
\providecommand \EOS [0]{\spacefactor3000\relax}%
\providecommand \BibitemShut  [1]{\csname bibitem#1\endcsname}%
\let\auto@bib@innerbib\@empty
\bibitem [{\citenamefont {Bardeen}\ \emph {et~al.}(1957)\citenamefont
  {Bardeen}, \citenamefont {Cooper},\ and\ \citenamefont
  {Schrieffer}}]{PhysRev.108.1175}%
  \BibitemOpen
  \bibfield  {author} {\bibinfo {author} {\bibfnamefont {J.}~\bibnamefont
  {Bardeen}}, \bibinfo {author} {\bibfnamefont {L.~N.}\ \bibnamefont {Cooper}},
  \ and\ \bibinfo {author} {\bibfnamefont {J.~R.}\ \bibnamefont {Schrieffer}},\
  }\href {\doibase 10.1103/PhysRev.108.1175} {\bibfield  {journal} {\bibinfo
  {journal} {Phys. Rev.}\ }\textbf {\bibinfo {volume} {108}},\ \bibinfo {pages}
  {1175} (\bibinfo {year} {1957})}\BibitemShut {NoStop}%
\bibitem [{\citenamefont {Marsiglio}\ and\ \citenamefont
  {Carbotte}(2008)}]{Marsiglio2008}%
  \BibitemOpen
  \bibfield  {author} {\bibinfo {author} {\bibfnamefont {F.}~\bibnamefont
  {Marsiglio}}\ and\ \bibinfo {author} {\bibfnamefont {J.~P.}\ \bibnamefont
  {Carbotte}},\ }\enquote {\bibinfo {title} {Electron-phonon
  superconductivity},}\ \ (\bibinfo  {publisher} {Springer Berlin Heidelberg},\
  \bibinfo {address} {Berlin, Heidelberg},\ \bibinfo {year} {2008})\ pp.\
  \bibinfo {pages} {73--162}\BibitemShut {NoStop}%
\bibitem [{\citenamefont {Carlson}\ \emph {et~al.}(2004)\citenamefont
  {Carlson}, \citenamefont {Kivelson}, \citenamefont {Orgad},\ and\
  \citenamefont {Emery}}]{Carlson2004}%
  \BibitemOpen
  \bibfield  {author} {\bibinfo {author} {\bibfnamefont {E.~W.}\ \bibnamefont
  {Carlson}}, \bibinfo {author} {\bibfnamefont {S.~A.}\ \bibnamefont
  {Kivelson}}, \bibinfo {author} {\bibfnamefont {D.}~\bibnamefont {Orgad}}, \
  and\ \bibinfo {author} {\bibfnamefont {V.~J.}\ \bibnamefont {Emery}},\
  }\enquote {\bibinfo {title} {Concepts in high temperature
  superconductivity},}\ in\ \href {\doibase 10.1007/978-3-642-18914-2_6} {\emph
  {\bibinfo {booktitle} {The Physics of Superconductors: Vol. II.
  Superconductivity in Nanostructures, High-Tc and Novel Superconductors,
  Organic Superconductors}}},\ \bibinfo {editor} {edited by\ \bibinfo {editor}
  {\bibfnamefont {K.~H.}\ \bibnamefont {Bennemann}}\ and\ \bibinfo {editor}
  {\bibfnamefont {J.~B.}\ \bibnamefont {Ketterson}}}\ (\bibinfo  {publisher}
  {Springer Berlin Heidelberg},\ \bibinfo {address} {Berlin, Heidelberg},\
  \bibinfo {year} {2004})\ pp.\ \bibinfo {pages} {275--451}\BibitemShut
  {NoStop}%
\bibitem [{\citenamefont {Chakraverty}(1979)}]{ChakravertyBipolaron}%
  \BibitemOpen
  \bibfield  {author} {\bibinfo {author} {\bibfnamefont {B.~K.}\ \bibnamefont
  {Chakraverty}},\ }\href {\doibase 10.1051/jphyslet:0197900400509900}
  {\bibfield  {journal} {\bibinfo  {journal} {J. Phys. (Paris) Lett.}\ }\textbf
  {\bibinfo {volume} {40}},\ \bibinfo {pages} {99} (\bibinfo {year}
  {1979})}\BibitemShut {NoStop}%
\bibitem [{\citenamefont {Chakraverty}\ \emph {et~al.}(1987)\citenamefont
  {Chakraverty}, \citenamefont {Feinberg}, \citenamefont {Hang},\ and\
  \citenamefont {Avignon}}]{chakraverty1987squeezed}%
  \BibitemOpen
  \bibfield  {author} {\bibinfo {author} {\bibfnamefont {B.}~\bibnamefont
  {Chakraverty}}, \bibinfo {author} {\bibfnamefont {D.}~\bibnamefont
  {Feinberg}}, \bibinfo {author} {\bibfnamefont {Z.}~\bibnamefont {Hang}}, \
  and\ \bibinfo {author} {\bibfnamefont {M.}~\bibnamefont {Avignon}},\ }\href
  {\doibase 10.1016/0038-1098(87)90608-9} {\bibfield  {journal} {\bibinfo
  {journal} {Solid State Commun.}\ }\textbf {\bibinfo {volume} {64}},\ \bibinfo
  {pages} {1147} (\bibinfo {year} {1987})}\BibitemShut {NoStop}%
\bibitem [{\citenamefont {Bonc\ifmmode~\breve{}\else \u{}\fi{}a}\ \emph
  {et~al.}(2000)\citenamefont {Bonc\ifmmode~\breve{}\else \u{}\fi{}a},
  \citenamefont {Katras\ifmmode~\breve{}\else \u{}\fi{}nik},\ and\
  \citenamefont {Trugman}}]{PhysRevLett.84.3153}%
  \BibitemOpen
  \bibfield  {author} {\bibinfo {author} {\bibfnamefont {J.}~\bibnamefont
  {Bonc\ifmmode~\breve{}\else \u{}\fi{}a}}, \bibinfo {author} {\bibfnamefont
  {T.}~\bibnamefont {Katras\ifmmode~\breve{}\else \u{}\fi{}nik}}, \ and\
  \bibinfo {author} {\bibfnamefont {S.~A.}\ \bibnamefont {Trugman}},\ }\href
  {\doibase 10.1103/PhysRevLett.84.3153} {\bibfield  {journal} {\bibinfo
  {journal} {Phys. Rev. Lett.}\ }\textbf {\bibinfo {volume} {84}},\ \bibinfo
  {pages} {3153} (\bibinfo {year} {2000})}\BibitemShut {NoStop}%
\bibitem [{\citenamefont {Alexandrov}(2001)}]{alexandrov2001breakdown}%
  \BibitemOpen
  \bibfield  {author} {\bibinfo {author} {\bibfnamefont {A.~S.}\ \bibnamefont
  {Alexandrov}},\ }\href {\doibase 10.1209/epl/i2001-00492-x} {\bibfield
  {journal} {\bibinfo  {journal} {Europhys. Lett.}\ }\textbf {\bibinfo {volume}
  {56}},\ \bibinfo {pages} {92} (\bibinfo {year} {2001})}\BibitemShut {NoStop}%
\bibitem [{\citenamefont {Esterlis}\ \emph
  {et~al.}(2018{\natexlab{a}})\citenamefont {Esterlis}, \citenamefont
  {Nosarzewski}, \citenamefont {Huang}, \citenamefont {Moritz}, \citenamefont
  {Devereaux}, \citenamefont {Scalapino},\ and\ \citenamefont
  {Kivelson}}]{esterlis2018breakdown}%
  \BibitemOpen
  \bibfield  {author} {\bibinfo {author} {\bibfnamefont {I.}~\bibnamefont
  {Esterlis}}, \bibinfo {author} {\bibfnamefont {B.}~\bibnamefont
  {Nosarzewski}}, \bibinfo {author} {\bibfnamefont {E.~W.}\ \bibnamefont
  {Huang}}, \bibinfo {author} {\bibfnamefont {B.}~\bibnamefont {Moritz}},
  \bibinfo {author} {\bibfnamefont {T.~P.}\ \bibnamefont {Devereaux}}, \bibinfo
  {author} {\bibfnamefont {D.~J.}\ \bibnamefont {Scalapino}}, \ and\ \bibinfo
  {author} {\bibfnamefont {S.~A.}\ \bibnamefont {Kivelson}},\ }\href {\doibase
  10.1103/PhysRevB.97.140501} {\bibfield  {journal} {\bibinfo  {journal} {Phys.
  Rev. B}\ }\textbf {\bibinfo {volume} {97}},\ \bibinfo {pages} {140501}
  (\bibinfo {year} {2018}{\natexlab{a}})}\BibitemShut {NoStop}%
\bibitem [{\citenamefont {Esterlis}\ \emph {et~al.}(2019)\citenamefont
  {Esterlis}, \citenamefont {Kivelson},\ and\ \citenamefont
  {Scalapino}}]{esterlis2019pseudogap}%
  \BibitemOpen
  \bibfield  {author} {\bibinfo {author} {\bibfnamefont {I.}~\bibnamefont
  {Esterlis}}, \bibinfo {author} {\bibfnamefont {S.~A.}\ \bibnamefont
  {Kivelson}}, \ and\ \bibinfo {author} {\bibfnamefont {D.~J.}\ \bibnamefont
  {Scalapino}},\ }\href {\doibase 10.1103/PhysRevB.99.174516} {\bibfield
  {journal} {\bibinfo  {journal} {Phys. Rev. B}\ }\textbf {\bibinfo {volume}
  {99}},\ \bibinfo {pages} {174516} (\bibinfo {year} {2019})}\BibitemShut
  {NoStop}%
\bibitem [{\citenamefont {Han}\ \emph {et~al.}(2020)\citenamefont {Han},
  \citenamefont {Kivelson},\ and\ \citenamefont
  {Yao}}]{PhysRevLett.125.167001}%
  \BibitemOpen
  \bibfield  {author} {\bibinfo {author} {\bibfnamefont {Z.}~\bibnamefont
  {Han}}, \bibinfo {author} {\bibfnamefont {S.~A.}\ \bibnamefont {Kivelson}}, \
  and\ \bibinfo {author} {\bibfnamefont {H.}~\bibnamefont {Yao}},\ }\href
  {\doibase 10.1103/PhysRevLett.125.167001} {\bibfield  {journal} {\bibinfo
  {journal} {Phys. Rev. Lett.}\ }\textbf {\bibinfo {volume} {125}},\ \bibinfo
  {pages} {167001} (\bibinfo {year} {2020})}\BibitemShut {NoStop}%
\bibitem [{\citenamefont {Freericks}(1993)}]{PhysRevB.48.3881}%
  \BibitemOpen
  \bibfield  {author} {\bibinfo {author} {\bibfnamefont {J.~K.}\ \bibnamefont
  {Freericks}},\ }\href {\doibase 10.1103/PhysRevB.48.3881} {\bibfield
  {journal} {\bibinfo  {journal} {Phys. Rev. B}\ }\textbf {\bibinfo {volume}
  {48}},\ \bibinfo {pages} {3881} (\bibinfo {year} {1993})}\BibitemShut
  {NoStop}%
\bibitem [{\citenamefont {Moussa}\ and\ \citenamefont
  {Cohen}(2006)}]{CohenBounds}%
  \BibitemOpen
  \bibfield  {author} {\bibinfo {author} {\bibfnamefont {J.~E.}\ \bibnamefont
  {Moussa}}\ and\ \bibinfo {author} {\bibfnamefont {M.~L.}\ \bibnamefont
  {Cohen}},\ }\href {\doibase 10.1103/PhysRevB.74.094520} {\bibfield  {journal}
  {\bibinfo  {journal} {Phys. Rev. B}\ }\textbf {\bibinfo {volume} {74}},\
  \bibinfo {pages} {094520} (\bibinfo {year} {2006})}\BibitemShut {NoStop}%
\bibitem [{\citenamefont {Esterlis}\ \emph
  {et~al.}(2018{\natexlab{b}})\citenamefont {Esterlis}, \citenamefont
  {Kivelson},\ and\ \citenamefont {Scalapino}}]{esterlis2018bound}%
  \BibitemOpen
  \bibfield  {author} {\bibinfo {author} {\bibfnamefont {I.}~\bibnamefont
  {Esterlis}}, \bibinfo {author} {\bibfnamefont {S.~A.}\ \bibnamefont
  {Kivelson}}, \ and\ \bibinfo {author} {\bibfnamefont {D.~J.}\ \bibnamefont
  {Scalapino}},\ }\href {\doibase 10.1038/s41535-018-0133-0} {\bibfield
  {journal} {\bibinfo  {journal} {npj Quantum Mater.}\ }\textbf {\bibinfo
  {volume} {3}},\ \bibinfo {pages} {1} (\bibinfo {year}
  {2018}{\natexlab{b}})}\BibitemShut {NoStop}%
\bibitem [{\citenamefont {Hofmann}\ \emph
  {et~al.}(2022{\natexlab{a}})\citenamefont {Hofmann}, \citenamefont
  {Chowdhury}, \citenamefont {Kivelson},\ and\ \citenamefont
  {Berg}}]{hofmann2022boundTc}%
  \BibitemOpen
  \bibfield  {author} {\bibinfo {author} {\bibfnamefont {J.~S.}\ \bibnamefont
  {Hofmann}}, \bibinfo {author} {\bibfnamefont {D.}~\bibnamefont {Chowdhury}},
  \bibinfo {author} {\bibfnamefont {S.~A.}\ \bibnamefont {Kivelson}}, \ and\
  \bibinfo {author} {\bibfnamefont {E.}~\bibnamefont {Berg}},\ }\href {\doibase
  10.1038/s41535-022-00491-1} {\bibfield  {journal} {\bibinfo  {journal} {npj
  Quantum Mater.}\ }\textbf {\bibinfo {volume} {7}},\ \bibinfo {pages} {83}
  (\bibinfo {year} {2022}{\natexlab{a}})}\BibitemShut {NoStop}%
\bibitem [{\citenamefont {Chubukov}\ \emph {et~al.}(2020)\citenamefont
  {Chubukov}, \citenamefont {Abanov}, \citenamefont {Esterlis},\ and\
  \citenamefont {Kivelson}}]{Chubukov}%
  \BibitemOpen
  \bibfield  {author} {\bibinfo {author} {\bibfnamefont {A.~V.}\ \bibnamefont
  {Chubukov}}, \bibinfo {author} {\bibfnamefont {A.}~\bibnamefont {Abanov}},
  \bibinfo {author} {\bibfnamefont {I.}~\bibnamefont {Esterlis}}, \ and\
  \bibinfo {author} {\bibfnamefont {S.~A.}\ \bibnamefont {Kivelson}},\ }\href
  {\doibase https://doi.org/10.1016/j.aop.2020.168190} {\bibfield  {journal}
  {\bibinfo  {journal} {Ann. of Phys.}\ }\textbf {\bibinfo {volume} {417}},\
  \bibinfo {pages} {168190} (\bibinfo {year} {2020})}\BibitemShut {NoStop}%
\bibitem [{\citenamefont {Hofmann}\ \emph
  {et~al.}(2022{\natexlab{b}})\citenamefont {Hofmann}, \citenamefont
  {Chowdhury}, \citenamefont {Kivelson},\ and\ \citenamefont
  {Berg}}]{hofmann2022heuristic}%
  \BibitemOpen
  \bibfield  {author} {\bibinfo {author} {\bibfnamefont {J.~S.}\ \bibnamefont
  {Hofmann}}, \bibinfo {author} {\bibfnamefont {D.}~\bibnamefont {Chowdhury}},
  \bibinfo {author} {\bibfnamefont {S.~A.}\ \bibnamefont {Kivelson}}, \ and\
  \bibinfo {author} {\bibfnamefont {E.}~\bibnamefont {Berg}},\ }\href@noop {}
  {\bibfield  {journal} {\bibinfo  {journal} {npj Quantum Materials}\ }\textbf
  {\bibinfo {volume} {7}},\ \bibinfo {pages} {83} (\bibinfo {year}
  {2022}{\natexlab{b}})}\BibitemShut {NoStop}%
\bibitem [{\citenamefont {Zhang}\ \emph
  {et~al.}(2023{\natexlab{a}})\citenamefont {Zhang}, \citenamefont {Sous},
  \citenamefont {Reichman}, \citenamefont {Berciu}, \citenamefont {Millis},
  \citenamefont {Prokof'ev},\ and\ \citenamefont
  {Svistunov}}]{zhang2023bipolaron}%
  \BibitemOpen
  \bibfield  {author} {\bibinfo {author} {\bibfnamefont {C.}~\bibnamefont
  {Zhang}}, \bibinfo {author} {\bibfnamefont {J.}~\bibnamefont {Sous}},
  \bibinfo {author} {\bibfnamefont {D.~R.}\ \bibnamefont {Reichman}}, \bibinfo
  {author} {\bibfnamefont {M.}~\bibnamefont {Berciu}}, \bibinfo {author}
  {\bibfnamefont {A.~J.}\ \bibnamefont {Millis}}, \bibinfo {author}
  {\bibfnamefont {N.~V.}\ \bibnamefont {Prokof'ev}}, \ and\ \bibinfo {author}
  {\bibfnamefont {B.~V.}\ \bibnamefont {Svistunov}},\ }\href {\doibase
  10.1103/PhysRevX.13.011010} {\bibfield  {journal} {\bibinfo  {journal} {Phys.
  Rev. X}\ }\textbf {\bibinfo {volume} {13}},\ \bibinfo {pages} {011010}
  (\bibinfo {year} {2023}{\natexlab{a}})}\BibitemShut {NoStop}%
\bibitem [{\citenamefont {Sous}\ \emph {et~al.}(2022)\citenamefont {Sous},
  \citenamefont {Zhang}, \citenamefont {Berciu}, \citenamefont {Reichman},
  \citenamefont {Svistunov}, \citenamefont {Prokof'ev},\ and\ \citenamefont
  {Millis}}]{SousCoulomb}%
  \BibitemOpen
  \bibfield  {author} {\bibinfo {author} {\bibfnamefont {J.}~\bibnamefont
  {Sous}}, \bibinfo {author} {\bibfnamefont {C.}~\bibnamefont {Zhang}},
  \bibinfo {author} {\bibfnamefont {M.}~\bibnamefont {Berciu}}, \bibinfo
  {author} {\bibfnamefont {D.~R.}\ \bibnamefont {Reichman}}, \bibinfo {author}
  {\bibfnamefont {B.~V.}\ \bibnamefont {Svistunov}}, \bibinfo {author}
  {\bibfnamefont {N.~V.}\ \bibnamefont {Prokof'ev}}, \ and\ \bibinfo {author}
  {\bibfnamefont {A.~J.}\ \bibnamefont {Millis}},\ }\href@noop {} {\bibfield
  {journal} {\bibinfo  {journal} {arXiv preprint arXiv:2210.14236}\ } (\bibinfo
  {year} {2022})}\BibitemShut {NoStop}%
\bibitem [{\citenamefont {Wang}\ \emph {et~al.}(2022)\citenamefont {Wang},
  \citenamefont {Jiang},\ and\ \citenamefont {Yao}}]{wang2022robust}%
  \BibitemOpen
  \bibfield  {author} {\bibinfo {author} {\bibfnamefont {H.-X.}\ \bibnamefont
  {Wang}}, \bibinfo {author} {\bibfnamefont {Y.-F.}\ \bibnamefont {Jiang}}, \
  and\ \bibinfo {author} {\bibfnamefont {H.}~\bibnamefont {Yao}},\ }\href@noop
  {} {\bibfield  {journal} {\bibinfo  {journal} {arXiv preprint
  arXiv:2211.09143}\ } (\bibinfo {year} {2022})}\BibitemShut {NoStop}%
\bibitem [{\citenamefont {Tanjaroon~Ly}\ \emph {et~al.}(2023)\citenamefont
  {Tanjaroon~Ly}, \citenamefont {Cohen-Stead}, \citenamefont
  {Malkaruge~Costa},\ and\ \citenamefont
  {Johnston}}]{tanjaroon2023comparative}%
  \BibitemOpen
  \bibfield  {author} {\bibinfo {author} {\bibfnamefont {A.}~\bibnamefont
  {Tanjaroon~Ly}}, \bibinfo {author} {\bibfnamefont {B.}~\bibnamefont
  {Cohen-Stead}}, \bibinfo {author} {\bibfnamefont {S.}~\bibnamefont
  {Malkaruge~Costa}}, \ and\ \bibinfo {author} {\bibfnamefont {S.}~\bibnamefont
  {Johnston}},\ }\href@noop {} {\bibfield  {journal} {\bibinfo  {journal}
  {arXiv preprint arXiv preprint arXiv:2307.10058}\ } (\bibinfo {year}
  {2023})}\BibitemShut {NoStop}%
\bibitem [{\citenamefont {Carbone}\ \emph {et~al.}(2021)\citenamefont
  {Carbone}, \citenamefont {Millis}, \citenamefont {Reichman},\ and\
  \citenamefont {Sous}}]{CarboneBond}%
  \BibitemOpen
  \bibfield  {author} {\bibinfo {author} {\bibfnamefont {M.~R.}\ \bibnamefont
  {Carbone}}, \bibinfo {author} {\bibfnamefont {A.~J.}\ \bibnamefont {Millis}},
  \bibinfo {author} {\bibfnamefont {D.~R.}\ \bibnamefont {Reichman}}, \ and\
  \bibinfo {author} {\bibfnamefont {J.}~\bibnamefont {Sous}},\ }\href {\doibase
  10.1103/PhysRevB.104.L140307} {\bibfield  {journal} {\bibinfo  {journal}
  {Phys. Rev. B}\ }\textbf {\bibinfo {volume} {104}},\ \bibinfo {pages}
  {L140307} (\bibinfo {year} {2021})}\BibitemShut {NoStop}%
\bibitem [{\citenamefont {Xing}\ \emph {et~al.}(2021)\citenamefont {Xing},
  \citenamefont {Chiu}, \citenamefont {Poletti}, \citenamefont {Scalettar},\
  and\ \citenamefont {Batrouni}}]{BoSSH2D}%
  \BibitemOpen
  \bibfield  {author} {\bibinfo {author} {\bibfnamefont {B.}~\bibnamefont
  {Xing}}, \bibinfo {author} {\bibfnamefont {W.-T.}\ \bibnamefont {Chiu}},
  \bibinfo {author} {\bibfnamefont {D.}~\bibnamefont {Poletti}}, \bibinfo
  {author} {\bibfnamefont {R.~T.}\ \bibnamefont {Scalettar}}, \ and\ \bibinfo
  {author} {\bibfnamefont {G.}~\bibnamefont {Batrouni}},\ }\href {\doibase
  10.1103/PhysRevLett.126.017601} {\bibfield  {journal} {\bibinfo  {journal}
  {Phys. Rev. Lett.}\ }\textbf {\bibinfo {volume} {126}},\ \bibinfo {pages}
  {017601} (\bibinfo {year} {2021})}\BibitemShut {NoStop}%
\bibitem [{\citenamefont {Cai}\ \emph {et~al.}(2021)\citenamefont {Cai},
  \citenamefont {Li},\ and\ \citenamefont {Yao}}]{PhysRevLett.127.247203}%
  \BibitemOpen
  \bibfield  {author} {\bibinfo {author} {\bibfnamefont {X.}~\bibnamefont
  {Cai}}, \bibinfo {author} {\bibfnamefont {Z.-X.}\ \bibnamefont {Li}}, \ and\
  \bibinfo {author} {\bibfnamefont {H.}~\bibnamefont {Yao}},\ }\href {\doibase
  10.1103/PhysRevLett.127.247203} {\bibfield  {journal} {\bibinfo  {journal}
  {Phys. Rev. Lett.}\ }\textbf {\bibinfo {volume} {127}},\ \bibinfo {pages}
  {247203} (\bibinfo {year} {2021})}\BibitemShut {NoStop}%
\bibitem [{\citenamefont {Feng}\ \emph {et~al.}(2021)\citenamefont {Feng},
  \citenamefont {Xing}, \citenamefont {Poletti}, \citenamefont {Scalettar},\
  and\ \citenamefont {Batrouni}}]{ScaletterSSHU}%
  \BibitemOpen
  \bibfield  {author} {\bibinfo {author} {\bibfnamefont {C.}~\bibnamefont
  {Feng}}, \bibinfo {author} {\bibfnamefont {B.}~\bibnamefont {Xing}}, \bibinfo
  {author} {\bibfnamefont {D.}~\bibnamefont {Poletti}}, \bibinfo {author}
  {\bibfnamefont {R.}~\bibnamefont {Scalettar}}, \ and\ \bibinfo {author}
  {\bibfnamefont {G.}~\bibnamefont {Batrouni}},\ }\href@noop {} {\bibfield
  {journal} {\bibinfo  {journal} {arXiv preprint arXiv:2109.09206}\ } (\bibinfo
  {year} {2021})}\BibitemShut {NoStop}%
\bibitem [{\citenamefont {Cai}\ \emph {et~al.}(2022)\citenamefont {Cai},
  \citenamefont {Li},\ and\ \citenamefont {Yao}}]{PhysRevB.106.L081115}%
  \BibitemOpen
  \bibfield  {author} {\bibinfo {author} {\bibfnamefont {X.}~\bibnamefont
  {Cai}}, \bibinfo {author} {\bibfnamefont {Z.-X.}\ \bibnamefont {Li}}, \ and\
  \bibinfo {author} {\bibfnamefont {H.}~\bibnamefont {Yao}},\ }\href {\doibase
  10.1103/PhysRevB.106.L081115} {\bibfield  {journal} {\bibinfo  {journal}
  {Phys. Rev. B}\ }\textbf {\bibinfo {volume} {106}},\ \bibinfo {pages}
  {L081115} (\bibinfo {year} {2022})}\BibitemShut {NoStop}%
\bibitem [{\citenamefont {G\"otz}\ \emph {et~al.}(2022)\citenamefont {G\"otz},
  \citenamefont {Beyl}, \citenamefont {Hohenadler},\ and\ \citenamefont
  {Assaad}}]{PhysRevB.105.085151}%
  \BibitemOpen
  \bibfield  {author} {\bibinfo {author} {\bibfnamefont {A.}~\bibnamefont
  {G\"otz}}, \bibinfo {author} {\bibfnamefont {S.}~\bibnamefont {Beyl}},
  \bibinfo {author} {\bibfnamefont {M.}~\bibnamefont {Hohenadler}}, \ and\
  \bibinfo {author} {\bibfnamefont {F.~F.}\ \bibnamefont {Assaad}},\ }\href
  {\doibase 10.1103/PhysRevB.105.085151} {\bibfield  {journal} {\bibinfo
  {journal} {Phys. Rev. B}\ }\textbf {\bibinfo {volume} {105}},\ \bibinfo
  {pages} {085151} (\bibinfo {year} {2022})}\BibitemShut {NoStop}%
\bibitem [{\citenamefont {Han}\ and\ \citenamefont
  {Kivelson}(2023)}]{PhysRevLett.130.186404}%
  \BibitemOpen
  \bibfield  {author} {\bibinfo {author} {\bibfnamefont {Z.}~\bibnamefont
  {Han}}\ and\ \bibinfo {author} {\bibfnamefont {S.~A.}\ \bibnamefont
  {Kivelson}},\ }\href {\doibase 10.1103/PhysRevLett.130.186404} {\bibfield
  {journal} {\bibinfo  {journal} {Phys. Rev. Lett.}\ }\textbf {\bibinfo
  {volume} {130}},\ \bibinfo {pages} {186404} (\bibinfo {year}
  {2023})}\BibitemShut {NoStop}%
\bibitem [{\citenamefont {G{\"o}tz}\ \emph {et~al.}(2023)\citenamefont
  {G{\"o}tz}, \citenamefont {Hohenadler},\ and\ \citenamefont
  {Assaad}}]{gotz2023phases}%
  \BibitemOpen
  \bibfield  {author} {\bibinfo {author} {\bibfnamefont {A.}~\bibnamefont
  {G{\"o}tz}}, \bibinfo {author} {\bibfnamefont {M.}~\bibnamefont
  {Hohenadler}}, \ and\ \bibinfo {author} {\bibfnamefont {F.~F.}\ \bibnamefont
  {Assaad}},\ }\href@noop {} {\bibfield  {journal} {\bibinfo  {journal} {arXiv
  preprint arXiv:2307.07613}\ } (\bibinfo {year} {2023})}\BibitemShut {NoStop}%
\bibitem [{\citenamefont {Costa}\ \emph {et~al.}(2023)\citenamefont {Costa},
  \citenamefont {Cohen-Stead}, \citenamefont {Tanjaroon~Ly}, \citenamefont
  {Neuhaus},\ and\ \citenamefont {Johnston}}]{costa2023comparative}%
  \BibitemOpen
  \bibfield  {author} {\bibinfo {author} {\bibfnamefont {S.~M.}\ \bibnamefont
  {Costa}}, \bibinfo {author} {\bibfnamefont {B.}~\bibnamefont {Cohen-Stead}},
  \bibinfo {author} {\bibfnamefont {A.}~\bibnamefont {Tanjaroon~Ly}}, \bibinfo
  {author} {\bibfnamefont {J.}~\bibnamefont {Neuhaus}}, \ and\ \bibinfo
  {author} {\bibfnamefont {S.}~\bibnamefont {Johnston}},\ }\href@noop {}
  {\bibfield  {journal} {\bibinfo  {journal} {arXiv preprint arXiv:2307.10058}\
  } (\bibinfo {year} {2023})}\BibitemShut {NoStop}%
\bibitem [{\citenamefont {Kim}\ \emph {et~al.}(2023)\citenamefont {Kim},
  \citenamefont {Han},\ and\ \citenamefont {Sous}}]{kim2023semi}%
  \BibitemOpen
  \bibfield  {author} {\bibinfo {author} {\bibfnamefont {K.-S.}\ \bibnamefont
  {Kim}}, \bibinfo {author} {\bibfnamefont {Z.}~\bibnamefont {Han}}, \ and\
  \bibinfo {author} {\bibfnamefont {J.}~\bibnamefont {Sous}},\ }\href@noop {}
  {\bibfield  {journal} {\bibinfo  {journal} {arXiv preprint arXiv:2308.01961}\
  } (\bibinfo {year} {2023})}\BibitemShut {NoStop}%
\bibitem [{\citenamefont {Ngai}(1974)}]{PhysRevLett.32.215}%
  \BibitemOpen
  \bibfield  {author} {\bibinfo {author} {\bibfnamefont {K.~L.}\ \bibnamefont
  {Ngai}},\ }\href {\doibase 10.1103/PhysRevLett.32.215} {\bibfield  {journal}
  {\bibinfo  {journal} {Phys. Rev. Lett.}\ }\textbf {\bibinfo {volume} {32}},\
  \bibinfo {pages} {215} (\bibinfo {year} {1974})}\BibitemShut {NoStop}%
\bibitem [{\citenamefont {Kuklov}(1989)}]{kuklov1989soliton}%
  \BibitemOpen
  \bibfield  {author} {\bibinfo {author} {\bibfnamefont {A.}~\bibnamefont
  {Kuklov}},\ }\href@noop {} {\bibfield  {journal} {\bibinfo  {journal}
  {Physics Letters A}\ }\textbf {\bibinfo {volume} {139}},\ \bibinfo {pages}
  {270} (\bibinfo {year} {1989})}\BibitemShut {NoStop}%
\bibitem [{\citenamefont {Entin-Wohlman}\ \emph {et~al.}(1983)\citenamefont
  {Entin-Wohlman}, \citenamefont {Gutfreund},\ and\ \citenamefont
  {Weger}}]{ENTINWOHLMAN19831}%
  \BibitemOpen
  \bibfield  {author} {\bibinfo {author} {\bibfnamefont {O.}~\bibnamefont
  {Entin-Wohlman}}, \bibinfo {author} {\bibfnamefont {H.}~\bibnamefont
  {Gutfreund}}, \ and\ \bibinfo {author} {\bibfnamefont {M.}~\bibnamefont
  {Weger}},\ }\href {\doibase https://doi.org/10.1016/0038-1098(83)90017-0}
  {\bibfield  {journal} {\bibinfo  {journal} {Solid State Communications}\
  }\textbf {\bibinfo {volume} {46}},\ \bibinfo {pages} {1} (\bibinfo {year}
  {1983})}\BibitemShut {NoStop}%
\bibitem [{\citenamefont {Riseborough}(1984)}]{RISEBOROUGH19841}%
  \BibitemOpen
  \bibfield  {author} {\bibinfo {author} {\bibfnamefont {P.~S.}\ \bibnamefont
  {Riseborough}},\ }\href {\doibase
  https://doi.org/10.1016/0003-4916(84)90183-0} {\bibfield  {journal} {\bibinfo
   {journal} {Annals of Physics}\ }\textbf {\bibinfo {volume} {153}},\ \bibinfo
  {pages} {1} (\bibinfo {year} {1984})}\BibitemShut {NoStop}%
\bibitem [{\citenamefont {Entin-Wohlman}\ \emph {et~al.}(1985)\citenamefont
  {Entin-Wohlman}, \citenamefont {Gutfreund},\ and\ \citenamefont
  {Weger}}]{Entin-Wohlman_1985}%
  \BibitemOpen
  \bibfield  {author} {\bibinfo {author} {\bibfnamefont {O.}~\bibnamefont
  {Entin-Wohlman}}, \bibinfo {author} {\bibfnamefont {H.}~\bibnamefont
  {Gutfreund}}, \ and\ \bibinfo {author} {\bibfnamefont {M.}~\bibnamefont
  {Weger}},\ }\href {\doibase 10.1088/0022-3719/18/3/001} {\bibfield  {journal}
  {\bibinfo  {journal} {Journal of Physics C: Solid State Physics}\ }\textbf
  {\bibinfo {volume} {18}},\ \bibinfo {pages} {L61} (\bibinfo {year}
  {1985})}\BibitemShut {NoStop}%
\bibitem [{\citenamefont {Hizhnyakov}(2010)}]{HIZHNYAKOV2010191}%
  \BibitemOpen
  \bibfield  {author} {\bibinfo {author} {\bibfnamefont {V.}~\bibnamefont
  {Hizhnyakov}},\ }\href {\doibase
  https://doi.org/10.1016/j.cplett.2010.05.017} {\bibfield  {journal} {\bibinfo
   {journal} {Chemical Physics Letters}\ }\textbf {\bibinfo {volume} {493}},\
  \bibinfo {pages} {191} (\bibinfo {year} {2010})}\BibitemShut {NoStop}%
\bibitem [{\citenamefont {Heid}(1992)}]{PhysRevB.45.5052}%
  \BibitemOpen
  \bibfield  {author} {\bibinfo {author} {\bibfnamefont {R.}~\bibnamefont
  {Heid}},\ }\href {\doibase 10.1103/PhysRevB.45.5052} {\bibfield  {journal}
  {\bibinfo  {journal} {Phys. Rev. B}\ }\textbf {\bibinfo {volume} {45}},\
  \bibinfo {pages} {5052} (\bibinfo {year} {1992})}\BibitemShut {NoStop}%
\bibitem [{\citenamefont {Matthew}\ and\ \citenamefont
  {Hart-Davis}(1968)}]{PhysRev.168.936}%
  \BibitemOpen
  \bibfield  {author} {\bibinfo {author} {\bibfnamefont {J.~A.~D.}\
  \bibnamefont {Matthew}}\ and\ \bibinfo {author} {\bibfnamefont
  {A.}~\bibnamefont {Hart-Davis}},\ }\href {\doibase 10.1103/PhysRev.168.936}
  {\bibfield  {journal} {\bibinfo  {journal} {Phys. Rev.}\ }\textbf {\bibinfo
  {volume} {168}},\ \bibinfo {pages} {936} (\bibinfo {year}
  {1968})}\BibitemShut {NoStop}%
\bibitem [{\citenamefont {Li}\ \emph {et~al.}(2015)\citenamefont {Li},
  \citenamefont {Nowadnick},\ and\ \citenamefont
  {Johnston}}]{PhysRevB.92.064301}%
  \BibitemOpen
  \bibfield  {author} {\bibinfo {author} {\bibfnamefont {S.}~\bibnamefont
  {Li}}, \bibinfo {author} {\bibfnamefont {E.~A.}\ \bibnamefont {Nowadnick}}, \
  and\ \bibinfo {author} {\bibfnamefont {S.}~\bibnamefont {Johnston}},\ }\href
  {\doibase 10.1103/PhysRevB.92.064301} {\bibfield  {journal} {\bibinfo
  {journal} {Phys. Rev. B}\ }\textbf {\bibinfo {volume} {92}},\ \bibinfo
  {pages} {064301} (\bibinfo {year} {2015})}\BibitemShut {NoStop}%
\bibitem [{\citenamefont {Gogolin}\ and\ \citenamefont
  {Ioselevich}(1991)}]{gogolin1991quantum}%
  \BibitemOpen
  \bibfield  {author} {\bibinfo {author} {\bibfnamefont {A.}~\bibnamefont
  {Gogolin}}\ and\ \bibinfo {author} {\bibfnamefont {A.}~\bibnamefont
  {Ioselevich}},\ }\href@noop {} {\bibfield  {journal} {\bibinfo  {journal}
  {JETP letters}\ }\textbf {\bibinfo {volume} {53}},\ \bibinfo {pages} {479}
  (\bibinfo {year} {1991})}\BibitemShut {NoStop}%
\bibitem [{\citenamefont {Adolphs}\ and\ \citenamefont
  {Berciu}(2014{\natexlab{a}})}]{PhysRevB.89.035122}%
  \BibitemOpen
  \bibfield  {author} {\bibinfo {author} {\bibfnamefont {C.~P.~J.}\
  \bibnamefont {Adolphs}}\ and\ \bibinfo {author} {\bibfnamefont
  {M.}~\bibnamefont {Berciu}},\ }\href {\doibase 10.1103/PhysRevB.89.035122}
  {\bibfield  {journal} {\bibinfo  {journal} {Phys. Rev. B}\ }\textbf {\bibinfo
  {volume} {89}},\ \bibinfo {pages} {035122} (\bibinfo {year}
  {2014}{\natexlab{a}})}\BibitemShut {NoStop}%
\bibitem [{\citenamefont {Adolphs}\ and\ \citenamefont
  {Berciu}(2014{\natexlab{b}})}]{PhysRevB.90.085149}%
  \BibitemOpen
  \bibfield  {author} {\bibinfo {author} {\bibfnamefont {C.~P.~J.}\
  \bibnamefont {Adolphs}}\ and\ \bibinfo {author} {\bibfnamefont
  {M.}~\bibnamefont {Berciu}},\ }\href {\doibase 10.1103/PhysRevB.90.085149}
  {\bibfield  {journal} {\bibinfo  {journal} {Phys. Rev. B}\ }\textbf {\bibinfo
  {volume} {90}},\ \bibinfo {pages} {085149} (\bibinfo {year}
  {2014}{\natexlab{b}})}\BibitemShut {NoStop}%
\bibitem [{\citenamefont {Mahan}(1997)}]{PhysRevB.56.8322}%
  \BibitemOpen
  \bibfield  {author} {\bibinfo {author} {\bibfnamefont {G.~D.}\ \bibnamefont
  {Mahan}},\ }\href {\doibase 10.1103/PhysRevB.56.8322} {\bibfield  {journal}
  {\bibinfo  {journal} {Phys. Rev. B}\ }\textbf {\bibinfo {volume} {56}},\
  \bibinfo {pages} {8322} (\bibinfo {year} {1997})}\BibitemShut {NoStop}%
\bibitem [{\citenamefont {Adolphs}\ and\ \citenamefont
  {Berciu}(2013)}]{adolphs2013going}%
  \BibitemOpen
  \bibfield  {author} {\bibinfo {author} {\bibfnamefont {C.~P.}\ \bibnamefont
  {Adolphs}}\ and\ \bibinfo {author} {\bibfnamefont {M.}~\bibnamefont
  {Berciu}},\ }\href@noop {} {\bibfield  {journal} {\bibinfo  {journal}
  {Europhysics Letters}\ }\textbf {\bibinfo {volume} {102}},\ \bibinfo {pages}
  {47003} (\bibinfo {year} {2013})}\BibitemShut {NoStop}%
\bibitem [{\citenamefont {Kiselov}\ and\ \citenamefont
  {Feigel'man}(2021{\natexlab{a}})}]{kiseliov}%
  \BibitemOpen
  \bibfield  {author} {\bibinfo {author} {\bibfnamefont {D.~E.}\ \bibnamefont
  {Kiselov}}\ and\ \bibinfo {author} {\bibfnamefont {M.~V.}\ \bibnamefont
  {Feigel'man}},\ }\href {\doibase 10.1103/PhysRevB.104.L220506} {\bibfield
  {journal} {\bibinfo  {journal} {Phys. Rev. B}\ }\textbf {\bibinfo {volume}
  {104}},\ \bibinfo {pages} {L220506} (\bibinfo {year}
  {2021}{\natexlab{a}})}\BibitemShut {NoStop}%
\bibitem [{\citenamefont {Volkov}\ \emph {et~al.}(2022)\citenamefont {Volkov},
  \citenamefont {Chandra},\ and\ \citenamefont
  {Coleman}}]{volkov2022superconductivity}%
  \BibitemOpen
  \bibfield  {author} {\bibinfo {author} {\bibfnamefont {P.~A.}\ \bibnamefont
  {Volkov}}, \bibinfo {author} {\bibfnamefont {P.}~\bibnamefont {Chandra}}, \
  and\ \bibinfo {author} {\bibfnamefont {P.}~\bibnamefont {Coleman}},\ }\href
  {\doibase 10.1038/s41467-022-32303-2} {\bibfield  {journal} {\bibinfo
  {journal} {Nature communications}\ }\textbf {\bibinfo {volume} {13}},\
  \bibinfo {pages} {4599} (\bibinfo {year} {2022})}\BibitemShut {NoStop}%
\bibitem [{\citenamefont {Yildirim}\ \emph {et~al.}(2001)\citenamefont
  {Yildirim}, \citenamefont {G\"ulseren}, \citenamefont {Lynn}, \citenamefont
  {Brown}, \citenamefont {Udovic}, \citenamefont {Huang}, \citenamefont
  {Rogado}, \citenamefont {Regan}, \citenamefont {Hayward}, \citenamefont
  {Slusky}, \citenamefont {He}, \citenamefont {Haas}, \citenamefont {Khalifah},
  \citenamefont {Inumaru},\ and\ \citenamefont {Cava}}]{PhysRevLett.87.037001}%
  \BibitemOpen
  \bibfield  {author} {\bibinfo {author} {\bibfnamefont {T.}~\bibnamefont
  {Yildirim}}, \bibinfo {author} {\bibfnamefont {O.}~\bibnamefont
  {G\"ulseren}}, \bibinfo {author} {\bibfnamefont {J.~W.}\ \bibnamefont
  {Lynn}}, \bibinfo {author} {\bibfnamefont {C.~M.}\ \bibnamefont {Brown}},
  \bibinfo {author} {\bibfnamefont {T.~J.}\ \bibnamefont {Udovic}}, \bibinfo
  {author} {\bibfnamefont {Q.}~\bibnamefont {Huang}}, \bibinfo {author}
  {\bibfnamefont {N.}~\bibnamefont {Rogado}}, \bibinfo {author} {\bibfnamefont
  {K.~A.}\ \bibnamefont {Regan}}, \bibinfo {author} {\bibfnamefont {M.~A.}\
  \bibnamefont {Hayward}}, \bibinfo {author} {\bibfnamefont {J.~S.}\
  \bibnamefont {Slusky}}, \bibinfo {author} {\bibfnamefont {T.}~\bibnamefont
  {He}}, \bibinfo {author} {\bibfnamefont {M.~K.}\ \bibnamefont {Haas}},
  \bibinfo {author} {\bibfnamefont {P.}~\bibnamefont {Khalifah}}, \bibinfo
  {author} {\bibfnamefont {K.}~\bibnamefont {Inumaru}}, \ and\ \bibinfo
  {author} {\bibfnamefont {R.~J.}\ \bibnamefont {Cava}},\ }\href {\doibase
  10.1103/PhysRevLett.87.037001} {\bibfield  {journal} {\bibinfo  {journal}
  {Phys. Rev. Lett.}\ }\textbf {\bibinfo {volume} {87}},\ \bibinfo {pages}
  {037001} (\bibinfo {year} {2001})}\BibitemShut {NoStop}%
\bibitem [{\citenamefont {Kiselov}\ and\ \citenamefont
  {Feigel'man}(2021{\natexlab{b}})}]{PhysRevB.104.L220506}%
  \BibitemOpen
  \bibfield  {author} {\bibinfo {author} {\bibfnamefont {D.~E.}\ \bibnamefont
  {Kiselov}}\ and\ \bibinfo {author} {\bibfnamefont {M.~V.}\ \bibnamefont
  {Feigel'man}},\ }\href {\doibase 10.1103/PhysRevB.104.L220506} {\bibfield
  {journal} {\bibinfo  {journal} {Phys. Rev. B}\ }\textbf {\bibinfo {volume}
  {104}},\ \bibinfo {pages} {L220506} (\bibinfo {year}
  {2021}{\natexlab{b}})}\BibitemShut {NoStop}%
\bibitem [{\citenamefont {van~der Marel}\ \emph {et~al.}(2019)\citenamefont
  {van~der Marel}, \citenamefont {Barantani},\ and\ \citenamefont
  {Rischau}}]{PhysRevResearch.1.013003}%
  \BibitemOpen
  \bibfield  {author} {\bibinfo {author} {\bibfnamefont {D.}~\bibnamefont
  {van~der Marel}}, \bibinfo {author} {\bibfnamefont {F.}~\bibnamefont
  {Barantani}}, \ and\ \bibinfo {author} {\bibfnamefont {C.~W.}\ \bibnamefont
  {Rischau}},\ }\href {\doibase 10.1103/PhysRevResearch.1.013003} {\bibfield
  {journal} {\bibinfo  {journal} {Phys. Rev. Res.}\ }\textbf {\bibinfo {volume}
  {1}},\ \bibinfo {pages} {013003} (\bibinfo {year} {2019})}\BibitemShut
  {NoStop}%
\bibitem [{\citenamefont {Kumar}\ \emph {et~al.}(2021)\citenamefont {Kumar},
  \citenamefont {Yudson},\ and\ \citenamefont {Maslov}}]{kumar2021}%
  \BibitemOpen
  \bibfield  {author} {\bibinfo {author} {\bibfnamefont {A.}~\bibnamefont
  {Kumar}}, \bibinfo {author} {\bibfnamefont {V.~I.}\ \bibnamefont {Yudson}}, \
  and\ \bibinfo {author} {\bibfnamefont {D.~L.}\ \bibnamefont {Maslov}},\
  }\href {\doibase 10.1103/PhysRevLett.126.076601} {\bibfield  {journal}
  {\bibinfo  {journal} {Phys. Rev. Lett.}\ }\textbf {\bibinfo {volume} {126}},\
  \bibinfo {pages} {076601} (\bibinfo {year} {2021})}\BibitemShut {NoStop}%
\bibitem [{\citenamefont {Ragni}\ \emph {et~al.}(2023)\citenamefont {Ragni},
  \citenamefont {Hahn}, \citenamefont {Zhang}, \citenamefont {Prokof'ev},
  \citenamefont {Kuklov}, \citenamefont {Klimin}, \citenamefont {Houtput},
  \citenamefont {Svistunov}, \citenamefont {Tempere}, \citenamefont {Nagaosa},
  \citenamefont {Franchini},\ and\ \citenamefont
  {Mishchenko}}]{PhysRevB.107.L121109}%
  \BibitemOpen
  \bibfield  {author} {\bibinfo {author} {\bibfnamefont {S.}~\bibnamefont
  {Ragni}}, \bibinfo {author} {\bibfnamefont {T.}~\bibnamefont {Hahn}},
  \bibinfo {author} {\bibfnamefont {Z.}~\bibnamefont {Zhang}}, \bibinfo
  {author} {\bibfnamefont {N.}~\bibnamefont {Prokof'ev}}, \bibinfo {author}
  {\bibfnamefont {A.}~\bibnamefont {Kuklov}}, \bibinfo {author} {\bibfnamefont
  {S.}~\bibnamefont {Klimin}}, \bibinfo {author} {\bibfnamefont
  {M.}~\bibnamefont {Houtput}}, \bibinfo {author} {\bibfnamefont
  {B.}~\bibnamefont {Svistunov}}, \bibinfo {author} {\bibfnamefont
  {J.}~\bibnamefont {Tempere}}, \bibinfo {author} {\bibfnamefont
  {N.}~\bibnamefont {Nagaosa}}, \bibinfo {author} {\bibfnamefont
  {C.}~\bibnamefont {Franchini}}, \ and\ \bibinfo {author} {\bibfnamefont
  {A.~S.}\ \bibnamefont {Mishchenko}},\ }\href {\doibase
  10.1103/PhysRevB.107.L121109} {\bibfield  {journal} {\bibinfo  {journal}
  {Phys. Rev. B}\ }\textbf {\bibinfo {volume} {107}},\ \bibinfo {pages}
  {L121109} (\bibinfo {year} {2023})}\BibitemShut {NoStop}%
\bibitem [{\citenamefont {Zhang}\ \emph
  {et~al.}(2023{\natexlab{b}})\citenamefont {Zhang}, \citenamefont {Kuklov},
  \citenamefont {Prokof'ev},\ and\ \citenamefont
  {Svistunov}}]{zhang2023soliton}%
  \BibitemOpen
  \bibfield  {author} {\bibinfo {author} {\bibfnamefont {Z.}~\bibnamefont
  {Zhang}}, \bibinfo {author} {\bibfnamefont {A.}~\bibnamefont {Kuklov}},
  \bibinfo {author} {\bibfnamefont {N.}~\bibnamefont {Prokof'ev}}, \ and\
  \bibinfo {author} {\bibfnamefont {B.}~\bibnamefont {Svistunov}},\ }\href@noop
  {} {\bibfield  {journal} {\bibinfo  {journal} {arXiv preprint
  arXiv:2309.10669}\ } (\bibinfo {year} {2023}{\natexlab{b}})}\BibitemShut
  {NoStop}%
\bibitem [{\citenamefont {Holstein}(1959)}]{holstein1959studies}%
  \BibitemOpen
  \bibfield  {author} {\bibinfo {author} {\bibfnamefont {T.}~\bibnamefont
  {Holstein}},\ }\href@noop {} {\bibfield  {journal} {\bibinfo  {journal}
  {Annals of physics}\ }\textbf {\bibinfo {volume} {8}},\ \bibinfo {pages}
  {325} (\bibinfo {year} {1959})}\BibitemShut {NoStop}%
\bibitem [{\citenamefont {Ku}\ \emph {et~al.}(2002)\citenamefont {Ku},
  \citenamefont {Trugman},\ and\ \citenamefont {Bon\ifmmode~\check{c}\else
  \v{c}\fi{}a}}]{PhysRevB.65.174306}%
  \BibitemOpen
  \bibfield  {author} {\bibinfo {author} {\bibfnamefont {L.-C.}\ \bibnamefont
  {Ku}}, \bibinfo {author} {\bibfnamefont {S.~A.}\ \bibnamefont {Trugman}}, \
  and\ \bibinfo {author} {\bibfnamefont {J.}~\bibnamefont
  {Bon\ifmmode~\check{c}\else \v{c}\fi{}a}},\ }\href {\doibase
  10.1103/PhysRevB.65.174306} {\bibfield  {journal} {\bibinfo  {journal} {Phys.
  Rev. B}\ }\textbf {\bibinfo {volume} {65}},\ \bibinfo {pages} {174306}
  (\bibinfo {year} {2002})}\BibitemShut {NoStop}%
\bibitem [{\citenamefont {Bon\ifmmode~\check{c}\else \v{c}\fi{}a}\ \emph
  {et~al.}(1999)\citenamefont {Bon\ifmmode~\check{c}\else \v{c}\fi{}a},
  \citenamefont {Trugman},\ and\ \citenamefont {Batisti\ifmmode~\acute{c}\else
  \'{c}\fi{}}}]{PhysRevB.60.1633}%
  \BibitemOpen
  \bibfield  {author} {\bibinfo {author} {\bibfnamefont {J.}~\bibnamefont
  {Bon\ifmmode~\check{c}\else \v{c}\fi{}a}}, \bibinfo {author} {\bibfnamefont
  {S.~A.}\ \bibnamefont {Trugman}}, \ and\ \bibinfo {author} {\bibfnamefont
  {I.}~\bibnamefont {Batisti\ifmmode~\acute{c}\else \'{c}\fi{}}},\ }\href
  {\doibase 10.1103/PhysRevB.60.1633} {\bibfield  {journal} {\bibinfo
  {journal} {Phys. Rev. B}\ }\textbf {\bibinfo {volume} {60}},\ \bibinfo
  {pages} {1633} (\bibinfo {year} {1999})}\BibitemShut {NoStop}%
\bibitem [{\citenamefont {Aubry}\ \emph {et~al.}(1992)\citenamefont {Aubry},
  \citenamefont {Abramovici},\ and\ \citenamefont
  {Raimbault}}]{aubry1992chaotic}%
  \BibitemOpen
  \bibfield  {author} {\bibinfo {author} {\bibfnamefont {S.}~\bibnamefont
  {Aubry}}, \bibinfo {author} {\bibfnamefont {G.}~\bibnamefont {Abramovici}}, \
  and\ \bibinfo {author} {\bibfnamefont {J.~L.}\ \bibnamefont {Raimbault}},\
  }\href@noop {} {\bibfield  {journal} {\bibinfo  {journal} {Journal of
  statistical physics}\ }\textbf {\bibinfo {volume} {67}},\ \bibinfo {pages}
  {675} (\bibinfo {year} {1992})}\BibitemShut {NoStop}%
\bibitem [{\citenamefont {Alexandrov}\ and\ \citenamefont
  {Devreese}(2010)}]{alexandrov2010advances}%
  \BibitemOpen
  \bibfield  {author} {\bibinfo {author} {\bibfnamefont {A.~S.}\ \bibnamefont
  {Alexandrov}}\ and\ \bibinfo {author} {\bibfnamefont {J.~T.}\ \bibnamefont
  {Devreese}},\ }\href@noop {} {\emph {\bibinfo {title} {Advances in polaron
  physics}}},\ Vol.\ \bibinfo {volume} {159}\ (\bibinfo  {publisher}
  {Springer},\ \bibinfo {year} {2010})\BibitemShut {NoStop}%
\bibitem [{\citenamefont {Trugman}\ and\ \citenamefont
  {Bon{\v{c}}a}(1999)}]{trugman1999polaron}%
  \BibitemOpen
  \bibfield  {author} {\bibinfo {author} {\bibfnamefont {S.}~\bibnamefont
  {Trugman}}\ and\ \bibinfo {author} {\bibfnamefont {J.}~\bibnamefont
  {Bon{\v{c}}a}},\ }\href@noop {} {\bibfield  {journal} {\bibinfo  {journal}
  {Journal of superconductivity}\ }\textbf {\bibinfo {volume} {12}},\ \bibinfo
  {pages} {221} (\bibinfo {year} {1999})}\BibitemShut {NoStop}%
\bibitem [{\citenamefont {Alexandrov}(2000)}]{PhysRevB.61.12315}%
  \BibitemOpen
  \bibfield  {author} {\bibinfo {author} {\bibfnamefont {A.~S.}\ \bibnamefont
  {Alexandrov}},\ }\href {\doibase 10.1103/PhysRevB.61.12315} {\bibfield
  {journal} {\bibinfo  {journal} {Phys. Rev. B}\ }\textbf {\bibinfo {volume}
  {61}},\ \bibinfo {pages} {12315} (\bibinfo {year} {2000})}\BibitemShut
  {NoStop}%
\bibitem [{\citenamefont {Ku}\ and\ \citenamefont
  {Trugman}(2007)}]{PhysRevB.75.014307}%
  \BibitemOpen
  \bibfield  {author} {\bibinfo {author} {\bibfnamefont {L.-C.}\ \bibnamefont
  {Ku}}\ and\ \bibinfo {author} {\bibfnamefont {S.~A.}\ \bibnamefont
  {Trugman}},\ }\href {\doibase 10.1103/PhysRevB.75.014307} {\bibfield
  {journal} {\bibinfo  {journal} {Phys. Rev. B}\ }\textbf {\bibinfo {volume}
  {75}},\ \bibinfo {pages} {014307} (\bibinfo {year} {2007})}\BibitemShut
  {NoStop}%
\bibitem [{\citenamefont {Bon\ifmmode~\check{c}\else \v{c}\fi{}a}\ \emph
  {et~al.}(2019)\citenamefont {Bon\ifmmode~\check{c}\else \v{c}\fi{}a},
  \citenamefont {Trugman},\ and\ \citenamefont {Berciu}}]{PhysRevB.100.094307}%
  \BibitemOpen
  \bibfield  {author} {\bibinfo {author} {\bibfnamefont {J.}~\bibnamefont
  {Bon\ifmmode~\check{c}\else \v{c}\fi{}a}}, \bibinfo {author} {\bibfnamefont
  {S.~A.}\ \bibnamefont {Trugman}}, \ and\ \bibinfo {author} {\bibfnamefont
  {M.}~\bibnamefont {Berciu}},\ }\href {\doibase 10.1103/PhysRevB.100.094307}
  {\bibfield  {journal} {\bibinfo  {journal} {Phys. Rev. B}\ }\textbf {\bibinfo
  {volume} {100}},\ \bibinfo {pages} {094307} (\bibinfo {year}
  {2019})}\BibitemShut {NoStop}%
\bibitem [{\citenamefont {Kalosakas}\ \emph {et~al.}(1998)\citenamefont
  {Kalosakas}, \citenamefont {Aubry},\ and\ \citenamefont
  {Tsironis}}]{PhysRevB.58.3094}%
  \BibitemOpen
  \bibfield  {author} {\bibinfo {author} {\bibfnamefont {G.}~\bibnamefont
  {Kalosakas}}, \bibinfo {author} {\bibfnamefont {S.}~\bibnamefont {Aubry}}, \
  and\ \bibinfo {author} {\bibfnamefont {G.~P.}\ \bibnamefont {Tsironis}},\
  }\href {\doibase 10.1103/PhysRevB.58.3094} {\bibfield  {journal} {\bibinfo
  {journal} {Phys. Rev. B}\ }\textbf {\bibinfo {volume} {58}},\ \bibinfo
  {pages} {3094} (\bibinfo {year} {1998})}\BibitemShut {NoStop}%
\bibitem [{\citenamefont {Bari{\v{s}}i{\'c}}\ and\ \citenamefont
  {Bari{\v{s}}i{\'c}}(2006)}]{barivsic2006quantum}%
  \BibitemOpen
  \bibfield  {author} {\bibinfo {author} {\bibfnamefont {O.-S.}\ \bibnamefont
  {Bari{\v{s}}i{\'c}}}\ and\ \bibinfo {author} {\bibfnamefont {S.}~\bibnamefont
  {Bari{\v{s}}i{\'c}}},\ }\href@noop {} {\bibfield  {journal} {\bibinfo
  {journal} {The European Physical Journal B-Condensed Matter and Complex
  Systems}\ }\textbf {\bibinfo {volume} {54}},\ \bibinfo {pages} {1} (\bibinfo
  {year} {2006})}\BibitemShut {NoStop}%
\bibitem [{\citenamefont {Alexandrov}\ and\ \citenamefont
  {Kornilovitch}(1999)}]{PhysRevLett.82.807}%
  \BibitemOpen
  \bibfield  {author} {\bibinfo {author} {\bibfnamefont {A.~S.}\ \bibnamefont
  {Alexandrov}}\ and\ \bibinfo {author} {\bibfnamefont {P.~E.}\ \bibnamefont
  {Kornilovitch}},\ }\href {\doibase 10.1103/PhysRevLett.82.807} {\bibfield
  {journal} {\bibinfo  {journal} {Phys. Rev. Lett.}\ }\textbf {\bibinfo
  {volume} {82}},\ \bibinfo {pages} {807} (\bibinfo {year} {1999})}\BibitemShut
  {NoStop}%
\bibitem [{\citenamefont {Jansen}\ \emph {et~al.}(2022)\citenamefont {Jansen},
  \citenamefont {Bon\ifmmode~\check{c}\else \v{c}\fi{}a},\ and\ \citenamefont
  {Heidrich-Meisner}}]{PhysRevB.106.155129}%
  \BibitemOpen
  \bibfield  {author} {\bibinfo {author} {\bibfnamefont {D.}~\bibnamefont
  {Jansen}}, \bibinfo {author} {\bibfnamefont {J.}~\bibnamefont
  {Bon\ifmmode~\check{c}\else \v{c}\fi{}a}}, \ and\ \bibinfo {author}
  {\bibfnamefont {F.}~\bibnamefont {Heidrich-Meisner}},\ }\href {\doibase
  10.1103/PhysRevB.106.155129} {\bibfield  {journal} {\bibinfo  {journal}
  {Phys. Rev. B}\ }\textbf {\bibinfo {volume} {106}},\ \bibinfo {pages}
  {155129} (\bibinfo {year} {2022})}\BibitemShut {NoStop}%
\bibitem [{\citenamefont {Fehske}\ and\ \citenamefont
  {Trugman}(2007)}]{fehske2007numerical}%
  \BibitemOpen
  \bibfield  {author} {\bibinfo {author} {\bibfnamefont {H.}~\bibnamefont
  {Fehske}}\ and\ \bibinfo {author} {\bibfnamefont {S.~A.}\ \bibnamefont
  {Trugman}},\ }\enquote {\bibinfo {title} {Numerical solution of the holstein
  polaron problem},}\ in\ \href {\doibase 10.1007/978-1-4020-6348-0_10} {\emph
  {\bibinfo {booktitle} {Polarons in Advanced Materials}}},\ \bibinfo {editor}
  {edited by\ \bibinfo {editor} {\bibfnamefont {A.~S.}\ \bibnamefont
  {Alexandrov}}}\ (\bibinfo  {publisher} {Springer Netherlands},\ \bibinfo
  {address} {Dordrecht},\ \bibinfo {year} {2007})\ pp.\ \bibinfo {pages}
  {393--461}\BibitemShut {NoStop}%
\bibitem [{\citenamefont {Zhao}\ \emph {et~al.}(2023)\citenamefont {Zhao},
  \citenamefont {Han}, \citenamefont {Kivelson},\ and\ \citenamefont
  {Esterlis}}]{PhysRevB.107.075142}%
  \BibitemOpen
  \bibfield  {author} {\bibinfo {author} {\bibfnamefont {S.}~\bibnamefont
  {Zhao}}, \bibinfo {author} {\bibfnamefont {Z.}~\bibnamefont {Han}}, \bibinfo
  {author} {\bibfnamefont {S.~A.}\ \bibnamefont {Kivelson}}, \ and\ \bibinfo
  {author} {\bibfnamefont {I.}~\bibnamefont {Esterlis}},\ }\href {\doibase
  10.1103/PhysRevB.107.075142} {\bibfield  {journal} {\bibinfo  {journal}
  {Phys. Rev. B}\ }\textbf {\bibinfo {volume} {107}},\ \bibinfo {pages}
  {075142} (\bibinfo {year} {2023})}\BibitemShut {NoStop}%
\bibitem [{\citenamefont {Bari{\v{s}}i{\'c}}\ and\ \citenamefont
  {Bari{\v{s}}i{\'c}}(2008)}]{barivsic2008phase}%
  \BibitemOpen
  \bibfield  {author} {\bibinfo {author} {\bibfnamefont {O.~S.}\ \bibnamefont
  {Bari{\v{s}}i{\'c}}}\ and\ \bibinfo {author} {\bibfnamefont {S.}~\bibnamefont
  {Bari{\v{s}}i{\'c}}},\ }\href@noop {} {\bibfield  {journal} {\bibinfo
  {journal} {The European Physical Journal B}\ }\textbf {\bibinfo {volume}
  {64}},\ \bibinfo {pages} {1} (\bibinfo {year} {2008})}\BibitemShut {NoStop}%
\bibitem [{\citenamefont {Gole\ifmmode~\check{z}\else \v{z}\fi{}}\ \emph
  {et~al.}(2012)\citenamefont {Gole\ifmmode~\check{z}\else \v{z}\fi{}},
  \citenamefont {Bon\ifmmode~\check{c}\else \v{c}\fi{}a}, \citenamefont
  {Vidmar},\ and\ \citenamefont {Trugman}}]{PhysRevLett.109.236402}%
  \BibitemOpen
  \bibfield  {author} {\bibinfo {author} {\bibfnamefont {D.}~\bibnamefont
  {Gole\ifmmode~\check{z}\else \v{z}\fi{}}}, \bibinfo {author} {\bibfnamefont
  {J.}~\bibnamefont {Bon\ifmmode~\check{c}\else \v{c}\fi{}a}}, \bibinfo
  {author} {\bibfnamefont {L.}~\bibnamefont {Vidmar}}, \ and\ \bibinfo {author}
  {\bibfnamefont {S.~A.}\ \bibnamefont {Trugman}},\ }\href {\doibase
  10.1103/PhysRevLett.109.236402} {\bibfield  {journal} {\bibinfo  {journal}
  {Phys. Rev. Lett.}\ }\textbf {\bibinfo {volume} {109}},\ \bibinfo {pages}
  {236402} (\bibinfo {year} {2012})}\BibitemShut {NoStop}%
\bibitem [{Note1()}]{Note1}%
  \BibitemOpen
  \bibinfo {note} {A similar theory at {\protect \it finite} phonon densities
  has been proposed to explain certain light-induced transient pairings~\cite
  {kennes2017transient,sous2021phonon}.}\BibitemShut {Stop}%
\bibitem [{\citenamefont {Gor'kov}(2016)}]{gorkov2016}%
  \BibitemOpen
  \bibfield  {author} {\bibinfo {author} {\bibfnamefont {L.~P.}\ \bibnamefont
  {Gor'kov}},\ }\href {\doibase 10.1103/PhysRevB.93.054517} {\bibfield
  {journal} {\bibinfo  {journal} {Phys. Rev. B}\ }\textbf {\bibinfo {volume}
  {93}},\ \bibinfo {pages} {054517} (\bibinfo {year} {2016})}\BibitemShut
  {NoStop}%
\bibitem [{\citenamefont {Chubukov}\ \emph {et~al.}(2016)\citenamefont
  {Chubukov}, \citenamefont {Eremin},\ and\ \citenamefont
  {Efremov}}]{chubukov2016}%
  \BibitemOpen
  \bibfield  {author} {\bibinfo {author} {\bibfnamefont {A.~V.}\ \bibnamefont
  {Chubukov}}, \bibinfo {author} {\bibfnamefont {I.}~\bibnamefont {Eremin}}, \
  and\ \bibinfo {author} {\bibfnamefont {D.~V.}\ \bibnamefont {Efremov}},\
  }\href {\doibase 10.1103/PhysRevB.93.174516} {\bibfield  {journal} {\bibinfo
  {journal} {Phys. Rev. B}\ }\textbf {\bibinfo {volume} {93}},\ \bibinfo
  {pages} {174516} (\bibinfo {year} {2016})}\BibitemShut {NoStop}%
\bibitem [{Note2()}]{Note2}%
  \BibitemOpen
  \bibinfo {note} {See Supplemental Material [url] for the derivation of
  effective coefficients and the estimate of $g$ in for STO, and Refs.~\cite
  {vanderbilt2001,wannierrev,bauerle1980soft} therein.}\BibitemShut {Stop}%
\bibitem [{\citenamefont {Kohno}\ and\ \citenamefont
  {Takahashi}(1997)}]{PhysRevB.56.3212}%
  \BibitemOpen
  \bibfield  {author} {\bibinfo {author} {\bibfnamefont {M.}~\bibnamefont
  {Kohno}}\ and\ \bibinfo {author} {\bibfnamefont {M.}~\bibnamefont
  {Takahashi}},\ }\href {\doibase 10.1103/PhysRevB.56.3212} {\bibfield
  {journal} {\bibinfo  {journal} {Phys. Rev. B}\ }\textbf {\bibinfo {volume}
  {56}},\ \bibinfo {pages} {3212} (\bibinfo {year} {1997})}\BibitemShut
  {NoStop}%
\bibitem [{\citenamefont {Batrouni}\ and\ \citenamefont
  {Scalettar}(2000)}]{PhysRevLett.84.1599}%
  \BibitemOpen
  \bibfield  {author} {\bibinfo {author} {\bibfnamefont {G.~G.}\ \bibnamefont
  {Batrouni}}\ and\ \bibinfo {author} {\bibfnamefont {R.~T.}\ \bibnamefont
  {Scalettar}},\ }\href {\doibase 10.1103/PhysRevLett.84.1599} {\bibfield
  {journal} {\bibinfo  {journal} {Phys. Rev. Lett.}\ }\textbf {\bibinfo
  {volume} {84}},\ \bibinfo {pages} {1599} (\bibinfo {year}
  {2000})}\BibitemShut {NoStop}%
\bibitem [{\citenamefont {Schmid}\ \emph {et~al.}(2002)\citenamefont {Schmid},
  \citenamefont {Todo}, \citenamefont {Troyer},\ and\ \citenamefont
  {Dorneich}}]{PhysRevLett.88.167208}%
  \BibitemOpen
  \bibfield  {author} {\bibinfo {author} {\bibfnamefont {G.}~\bibnamefont
  {Schmid}}, \bibinfo {author} {\bibfnamefont {S.}~\bibnamefont {Todo}},
  \bibinfo {author} {\bibfnamefont {M.}~\bibnamefont {Troyer}}, \ and\ \bibinfo
  {author} {\bibfnamefont {A.}~\bibnamefont {Dorneich}},\ }\href {\doibase
  10.1103/PhysRevLett.88.167208} {\bibfield  {journal} {\bibinfo  {journal}
  {Phys. Rev. Lett.}\ }\textbf {\bibinfo {volume} {88}},\ \bibinfo {pages}
  {167208} (\bibinfo {year} {2002})}\BibitemShut {NoStop}%
\bibitem [{\citenamefont {Bonnes}\ and\ \citenamefont
  {Wessel}(2011)}]{PhysRevB.84.054510}%
  \BibitemOpen
  \bibfield  {author} {\bibinfo {author} {\bibfnamefont {L.}~\bibnamefont
  {Bonnes}}\ and\ \bibinfo {author} {\bibfnamefont {S.}~\bibnamefont
  {Wessel}},\ }\href {\doibase 10.1103/PhysRevB.84.054510} {\bibfield
  {journal} {\bibinfo  {journal} {Phys. Rev. B}\ }\textbf {\bibinfo {volume}
  {84}},\ \bibinfo {pages} {054510} (\bibinfo {year} {2011})}\BibitemShut
  {NoStop}%
\bibitem [{\citenamefont {Chen}\ \emph {et~al.}(2008)\citenamefont {Chen},
  \citenamefont {Melko}, \citenamefont {Wessel},\ and\ \citenamefont
  {Kao}}]{PhysRevB.77.014524}%
  \BibitemOpen
  \bibfield  {author} {\bibinfo {author} {\bibfnamefont {Y.-C.}\ \bibnamefont
  {Chen}}, \bibinfo {author} {\bibfnamefont {R.~G.}\ \bibnamefont {Melko}},
  \bibinfo {author} {\bibfnamefont {S.}~\bibnamefont {Wessel}}, \ and\ \bibinfo
  {author} {\bibfnamefont {Y.-J.}\ \bibnamefont {Kao}},\ }\href {\doibase
  10.1103/PhysRevB.77.014524} {\bibfield  {journal} {\bibinfo  {journal} {Phys.
  Rev. B}\ }\textbf {\bibinfo {volume} {77}},\ \bibinfo {pages} {014524}
  (\bibinfo {year} {2008})}\BibitemShut {NoStop}%
\bibitem [{\citenamefont {Melko}\ \emph {et~al.}(2005)\citenamefont {Melko},
  \citenamefont {Paramekanti}, \citenamefont {Burkov}, \citenamefont
  {Vishwanath}, \citenamefont {Sheng},\ and\ \citenamefont
  {Balents}}]{PhysRevLett.95.127207}%
  \BibitemOpen
  \bibfield  {author} {\bibinfo {author} {\bibfnamefont {R.~G.}\ \bibnamefont
  {Melko}}, \bibinfo {author} {\bibfnamefont {A.}~\bibnamefont {Paramekanti}},
  \bibinfo {author} {\bibfnamefont {A.~A.}\ \bibnamefont {Burkov}}, \bibinfo
  {author} {\bibfnamefont {A.}~\bibnamefont {Vishwanath}}, \bibinfo {author}
  {\bibfnamefont {D.~N.}\ \bibnamefont {Sheng}}, \ and\ \bibinfo {author}
  {\bibfnamefont {L.}~\bibnamefont {Balents}},\ }\href {\doibase
  10.1103/PhysRevLett.95.127207} {\bibfield  {journal} {\bibinfo  {journal}
  {Phys. Rev. Lett.}\ }\textbf {\bibinfo {volume} {95}},\ \bibinfo {pages}
  {127207} (\bibinfo {year} {2005})}\BibitemShut {NoStop}%
\bibitem [{\citenamefont {Wang}\ \emph {et~al.}(2009)\citenamefont {Wang},
  \citenamefont {Pollmann},\ and\ \citenamefont
  {Vishwanath}}]{PhysRevLett.102.017203}%
  \BibitemOpen
  \bibfield  {author} {\bibinfo {author} {\bibfnamefont {F.}~\bibnamefont
  {Wang}}, \bibinfo {author} {\bibfnamefont {F.}~\bibnamefont {Pollmann}}, \
  and\ \bibinfo {author} {\bibfnamefont {A.}~\bibnamefont {Vishwanath}},\
  }\href {\doibase 10.1103/PhysRevLett.102.017203} {\bibfield  {journal}
  {\bibinfo  {journal} {Phys. Rev. Lett.}\ }\textbf {\bibinfo {volume} {102}},\
  \bibinfo {pages} {017203} (\bibinfo {year} {2009})}\BibitemShut {NoStop}%
\bibitem [{\citenamefont {Wessel}(2007)}]{PhysRevB.75.174301}%
  \BibitemOpen
  \bibfield  {author} {\bibinfo {author} {\bibfnamefont {S.}~\bibnamefont
  {Wessel}},\ }\href {\doibase 10.1103/PhysRevB.75.174301} {\bibfield
  {journal} {\bibinfo  {journal} {Phys. Rev. B}\ }\textbf {\bibinfo {volume}
  {75}},\ \bibinfo {pages} {174301} (\bibinfo {year} {2007})}\BibitemShut
  {NoStop}%
\bibitem [{\citenamefont {Yamamoto}\ \emph {et~al.}(2014)\citenamefont
  {Yamamoto}, \citenamefont {Marmorini},\ and\ \citenamefont
  {Danshita}}]{PhysRevLett.112.127203}%
  \BibitemOpen
  \bibfield  {author} {\bibinfo {author} {\bibfnamefont {D.}~\bibnamefont
  {Yamamoto}}, \bibinfo {author} {\bibfnamefont {G.}~\bibnamefont {Marmorini}},
  \ and\ \bibinfo {author} {\bibfnamefont {I.}~\bibnamefont {Danshita}},\
  }\href {\doibase 10.1103/PhysRevLett.112.127203} {\bibfield  {journal}
  {\bibinfo  {journal} {Phys. Rev. Lett.}\ }\textbf {\bibinfo {volume} {112}},\
  \bibinfo {pages} {127203} (\bibinfo {year} {2014})}\BibitemShut {NoStop}%
\bibitem [{\citenamefont {Isakov}\ \emph {et~al.}(2006)\citenamefont {Isakov},
  \citenamefont {Wessel}, \citenamefont {Melko}, \citenamefont {Sengupta},\
  and\ \citenamefont {Kim}}]{PhysRevLett.97.147202}%
  \BibitemOpen
  \bibfield  {author} {\bibinfo {author} {\bibfnamefont {S.~V.}\ \bibnamefont
  {Isakov}}, \bibinfo {author} {\bibfnamefont {S.}~\bibnamefont {Wessel}},
  \bibinfo {author} {\bibfnamefont {R.~G.}\ \bibnamefont {Melko}}, \bibinfo
  {author} {\bibfnamefont {K.}~\bibnamefont {Sengupta}}, \ and\ \bibinfo
  {author} {\bibfnamefont {Y.~B.}\ \bibnamefont {Kim}},\ }\href {\doibase
  10.1103/PhysRevLett.97.147202} {\bibfield  {journal} {\bibinfo  {journal}
  {Phys. Rev. Lett.}\ }\textbf {\bibinfo {volume} {97}},\ \bibinfo {pages}
  {147202} (\bibinfo {year} {2006})}\BibitemShut {NoStop}%
\bibitem [{\citenamefont {Boninsegni}\ and\ \citenamefont
  {Prokof'ev}(2005)}]{PhysRevLett.95.237204}%
  \BibitemOpen
  \bibfield  {author} {\bibinfo {author} {\bibfnamefont {M.}~\bibnamefont
  {Boninsegni}}\ and\ \bibinfo {author} {\bibfnamefont {N.}~\bibnamefont
  {Prokof'ev}},\ }\href {\doibase 10.1103/PhysRevLett.95.237204} {\bibfield
  {journal} {\bibinfo  {journal} {Phys. Rev. Lett.}\ }\textbf {\bibinfo
  {volume} {95}},\ \bibinfo {pages} {237204} (\bibinfo {year}
  {2005})}\BibitemShut {NoStop}%
\bibitem [{\citenamefont {Sellmann}\ \emph {et~al.}(2015)\citenamefont
  {Sellmann}, \citenamefont {Zhang},\ and\ \citenamefont
  {Eggert}}]{PhysRevB.91.081104}%
  \BibitemOpen
  \bibfield  {author} {\bibinfo {author} {\bibfnamefont {D.}~\bibnamefont
  {Sellmann}}, \bibinfo {author} {\bibfnamefont {X.-F.}\ \bibnamefont {Zhang}},
  \ and\ \bibinfo {author} {\bibfnamefont {S.}~\bibnamefont {Eggert}},\ }\href
  {\doibase 10.1103/PhysRevB.91.081104} {\bibfield  {journal} {\bibinfo
  {journal} {Phys. Rev. B}\ }\textbf {\bibinfo {volume} {91}},\ \bibinfo
  {pages} {081104} (\bibinfo {year} {2015})}\BibitemShut {NoStop}%
\bibitem [{\citenamefont {Zhang}\ \emph {et~al.}(2011)\citenamefont {Zhang},
  \citenamefont {Dillenschneider}, \citenamefont {Yu},\ and\ \citenamefont
  {Eggert}}]{PhysRevB.84.174515}%
  \BibitemOpen
  \bibfield  {author} {\bibinfo {author} {\bibfnamefont {X.-F.}\ \bibnamefont
  {Zhang}}, \bibinfo {author} {\bibfnamefont {R.}~\bibnamefont
  {Dillenschneider}}, \bibinfo {author} {\bibfnamefont {Y.}~\bibnamefont {Yu}},
  \ and\ \bibinfo {author} {\bibfnamefont {S.}~\bibnamefont {Eggert}},\ }\href
  {\doibase 10.1103/PhysRevB.84.174515} {\bibfield  {journal} {\bibinfo
  {journal} {Phys. Rev. B}\ }\textbf {\bibinfo {volume} {84}},\ \bibinfo
  {pages} {174515} (\bibinfo {year} {2011})}\BibitemShut {NoStop}%
\bibitem [{\citenamefont {Mishra}\ \emph {et~al.}(2014)\citenamefont {Mishra},
  \citenamefont {Pai},\ and\ \citenamefont {Mukerjee}}]{PhysRevA.89.013615}%
  \BibitemOpen
  \bibfield  {author} {\bibinfo {author} {\bibfnamefont {T.}~\bibnamefont
  {Mishra}}, \bibinfo {author} {\bibfnamefont {R.~V.}\ \bibnamefont {Pai}}, \
  and\ \bibinfo {author} {\bibfnamefont {S.}~\bibnamefont {Mukerjee}},\ }\href
  {\doibase 10.1103/PhysRevA.89.013615} {\bibfield  {journal} {\bibinfo
  {journal} {Phys. Rev. A}\ }\textbf {\bibinfo {volume} {89}},\ \bibinfo
  {pages} {013615} (\bibinfo {year} {2014})}\BibitemShut {NoStop}%
\bibitem [{\citenamefont {Spevak}\ \emph {et~al.}(2021)\citenamefont {Spevak},
  \citenamefont {Panov},\ and\ \citenamefont {Moskvin}}]{spevak2021critical}%
  \BibitemOpen
  \bibfield  {author} {\bibinfo {author} {\bibfnamefont {E.~L.}\ \bibnamefont
  {Spevak}}, \bibinfo {author} {\bibfnamefont {Y.~D.}\ \bibnamefont {Panov}}, \
  and\ \bibinfo {author} {\bibfnamefont {A.~S.}\ \bibnamefont {Moskvin}},\
  }\href@noop {} {\bibfield  {journal} {\bibinfo  {journal} {Physics of the
  Solid State}\ }\textbf {\bibinfo {volume} {63}},\ \bibinfo {pages} {1546}
  (\bibinfo {year} {2021})}\BibitemShut {NoStop}%
\bibitem [{\citenamefont {Fisher}\ and\ \citenamefont
  {Hohenberg}(1988)}]{PhysRevB.37.4936}%
  \BibitemOpen
  \bibfield  {author} {\bibinfo {author} {\bibfnamefont {D.~S.}\ \bibnamefont
  {Fisher}}\ and\ \bibinfo {author} {\bibfnamefont {P.~C.}\ \bibnamefont
  {Hohenberg}},\ }\href {\doibase 10.1103/PhysRevB.37.4936} {\bibfield
  {journal} {\bibinfo  {journal} {Phys. Rev. B}\ }\textbf {\bibinfo {volume}
  {37}},\ \bibinfo {pages} {4936} (\bibinfo {year} {1988})}\BibitemShut
  {NoStop}%
\bibitem [{\citenamefont {Prokof'ev}\ \emph {et~al.}(2001)\citenamefont
  {Prokof'ev}, \citenamefont {Ruebenacker},\ and\ \citenamefont
  {Svistunov}}]{PhysRevLett.87.270402}%
  \BibitemOpen
  \bibfield  {author} {\bibinfo {author} {\bibfnamefont {N.}~\bibnamefont
  {Prokof'ev}}, \bibinfo {author} {\bibfnamefont {O.}~\bibnamefont
  {Ruebenacker}}, \ and\ \bibinfo {author} {\bibfnamefont {B.}~\bibnamefont
  {Svistunov}},\ }\href {\doibase 10.1103/PhysRevLett.87.270402} {\bibfield
  {journal} {\bibinfo  {journal} {Phys. Rev. Lett.}\ }\textbf {\bibinfo
  {volume} {87}},\ \bibinfo {pages} {270402} (\bibinfo {year}
  {2001})}\BibitemShut {NoStop}%
\bibitem [{\citenamefont {Pilati}\ \emph {et~al.}(2008)\citenamefont {Pilati},
  \citenamefont {Giorgini},\ and\ \citenamefont
  {Prokof'ev}}]{PhysRevLett.100.140405}%
  \BibitemOpen
  \bibfield  {author} {\bibinfo {author} {\bibfnamefont {S.}~\bibnamefont
  {Pilati}}, \bibinfo {author} {\bibfnamefont {S.}~\bibnamefont {Giorgini}}, \
  and\ \bibinfo {author} {\bibfnamefont {N.}~\bibnamefont {Prokof'ev}},\ }\href
  {\doibase 10.1103/PhysRevLett.100.140405} {\bibfield  {journal} {\bibinfo
  {journal} {Phys. Rev. Lett.}\ }\textbf {\bibinfo {volume} {100}},\ \bibinfo
  {pages} {140405} (\bibinfo {year} {2008})}\BibitemShut {NoStop}%
\bibitem [{\citenamefont {Zhang}\ \emph
  {et~al.}(2023{\natexlab{c}})\citenamefont {Zhang}, \citenamefont
  {Capogrosso-Sansone}, \citenamefont {Boninsegni}, \citenamefont {Prokof'ev},\
  and\ \citenamefont {Svistunov}}]{PhysRevLett.130.236001}%
  \BibitemOpen
  \bibfield  {author} {\bibinfo {author} {\bibfnamefont {C.}~\bibnamefont
  {Zhang}}, \bibinfo {author} {\bibfnamefont {B.}~\bibnamefont
  {Capogrosso-Sansone}}, \bibinfo {author} {\bibfnamefont {M.}~\bibnamefont
  {Boninsegni}}, \bibinfo {author} {\bibfnamefont {N.~V.}\ \bibnamefont
  {Prokof'ev}}, \ and\ \bibinfo {author} {\bibfnamefont {B.~V.}\ \bibnamefont
  {Svistunov}},\ }\href {\doibase 10.1103/PhysRevLett.130.236001} {\bibfield
  {journal} {\bibinfo  {journal} {Phys. Rev. Lett.}\ }\textbf {\bibinfo
  {volume} {130}},\ \bibinfo {pages} {236001} (\bibinfo {year}
  {2023}{\natexlab{c}})}\BibitemShut {NoStop}%
\bibitem [{\citenamefont {Kozii}\ and\ \citenamefont {Fu}(2015)}]{kozii2015}%
  \BibitemOpen
  \bibfield  {author} {\bibinfo {author} {\bibfnamefont {V.}~\bibnamefont
  {Kozii}}\ and\ \bibinfo {author} {\bibfnamefont {L.}~\bibnamefont {Fu}},\
  }\href {\doibase 10.1103/PhysRevLett.115.207002} {\bibfield  {journal}
  {\bibinfo  {journal} {Phys. Rev. Lett.}\ }\textbf {\bibinfo {volume} {115}},\
  \bibinfo {pages} {207002} (\bibinfo {year} {2015})}\BibitemShut {NoStop}%
\bibitem [{\citenamefont {Volkov}\ and\ \citenamefont
  {Chandra}(2020)}]{volkov2020}%
  \BibitemOpen
  \bibfield  {author} {\bibinfo {author} {\bibfnamefont {P.~A.}\ \bibnamefont
  {Volkov}}\ and\ \bibinfo {author} {\bibfnamefont {P.}~\bibnamefont
  {Chandra}},\ }\href {\doibase 10.1103/PhysRevLett.124.237601} {\bibfield
  {journal} {\bibinfo  {journal} {Phys. Rev. Lett.}\ }\textbf {\bibinfo
  {volume} {124}},\ \bibinfo {pages} {237601} (\bibinfo {year}
  {2020})}\BibitemShut {NoStop}%
\bibitem [{\citenamefont {Park}\ \emph {et~al.}(2021)\citenamefont {Park},
  \citenamefont {Cao}, \citenamefont {Watanabe}, \citenamefont {Taniguchi},\
  and\ \citenamefont {Jarillo-Herrero}}]{park2021tunable}%
  \BibitemOpen
  \bibfield  {author} {\bibinfo {author} {\bibfnamefont {J.~M.}\ \bibnamefont
  {Park}}, \bibinfo {author} {\bibfnamefont {Y.}~\bibnamefont {Cao}}, \bibinfo
  {author} {\bibfnamefont {K.}~\bibnamefont {Watanabe}}, \bibinfo {author}
  {\bibfnamefont {T.}~\bibnamefont {Taniguchi}}, \ and\ \bibinfo {author}
  {\bibfnamefont {P.}~\bibnamefont {Jarillo-Herrero}},\ }\href@noop {}
  {\bibfield  {journal} {\bibinfo  {journal} {Nature}\ }\textbf {\bibinfo
  {volume} {590}},\ \bibinfo {pages} {249} (\bibinfo {year}
  {2021})}\BibitemShut {NoStop}%
\bibitem [{\citenamefont {B{\"a}uerle}\ \emph {et~al.}(1980)\citenamefont
  {B{\"a}uerle}, \citenamefont {Wagner}, \citenamefont {W{\"o}hlecke},
  \citenamefont {Dorner},\ and\ \citenamefont
  {Kraxenberger}}]{bauerle1980soft}%
  \BibitemOpen
  \bibfield  {author} {\bibinfo {author} {\bibfnamefont {D.}~\bibnamefont
  {B{\"a}uerle}}, \bibinfo {author} {\bibfnamefont {D.}~\bibnamefont {Wagner}},
  \bibinfo {author} {\bibfnamefont {M.}~\bibnamefont {W{\"o}hlecke}}, \bibinfo
  {author} {\bibfnamefont {B.}~\bibnamefont {Dorner}}, \ and\ \bibinfo {author}
  {\bibfnamefont {H.}~\bibnamefont {Kraxenberger}},\ }\href@noop {} {\bibfield
  {journal} {\bibinfo  {journal} {Zeitschrift f{\"u}r Physik B Condensed
  Matter}\ }\textbf {\bibinfo {volume} {38}},\ \bibinfo {pages} {335} (\bibinfo
  {year} {1980})}\BibitemShut {NoStop}%
\bibitem [{Note3()}]{Note3}%
  \BibitemOpen
  \bibinfo {note} {We note that superconductivity originating from certain
  extended forms of bipolarons may still be stabilized when the Coulomb
  repulsion is stronger than $U_\protect \text {e-ph}$~\cite
  {PhysRevLett.84.3153,proville1999small,proville2000small,PROVILLE1998307,barivsic2012bipolarons,PhysRevLett.125.167001,PhysRevB.64.094507,PhysRevB.80.054301,PhysRevB.85.104520,hague2007superlight,PhysRevLett.98.037002}.}\BibitemShut
  {Stop}%
\bibitem [{\citenamefont {Forsythe}\ \emph {et~al.}(2018)\citenamefont
  {Forsythe}, \citenamefont {Zhou}, \citenamefont {Watanabe}, \citenamefont
  {Taniguchi}, \citenamefont {Pasupathy}, \citenamefont {Moon}, \citenamefont
  {Koshino}, \citenamefont {Kim},\ and\ \citenamefont
  {Dean}}]{forsythe2018band}%
  \BibitemOpen
  \bibfield  {author} {\bibinfo {author} {\bibfnamefont {C.}~\bibnamefont
  {Forsythe}}, \bibinfo {author} {\bibfnamefont {X.}~\bibnamefont {Zhou}},
  \bibinfo {author} {\bibfnamefont {K.}~\bibnamefont {Watanabe}}, \bibinfo
  {author} {\bibfnamefont {T.}~\bibnamefont {Taniguchi}}, \bibinfo {author}
  {\bibfnamefont {A.}~\bibnamefont {Pasupathy}}, \bibinfo {author}
  {\bibfnamefont {P.}~\bibnamefont {Moon}}, \bibinfo {author} {\bibfnamefont
  {M.}~\bibnamefont {Koshino}}, \bibinfo {author} {\bibfnamefont
  {P.}~\bibnamefont {Kim}}, \ and\ \bibinfo {author} {\bibfnamefont {C.~R.}\
  \bibnamefont {Dean}},\ }\href@noop {} {\bibfield  {journal} {\bibinfo
  {journal} {Nature nanotechnology}\ }\textbf {\bibinfo {volume} {13}},\
  \bibinfo {pages} {566} (\bibinfo {year} {2018})}\BibitemShut {NoStop}%
\bibitem [{\citenamefont {Liu}\ \emph {et~al.}(2021)\citenamefont {Liu},
  \citenamefont {Wang}, \citenamefont {Watanabe}, \citenamefont {Taniguchi},
  \citenamefont {Vafek},\ and\ \citenamefont {Li}}]{liu2021tuning}%
  \BibitemOpen
  \bibfield  {author} {\bibinfo {author} {\bibfnamefont {X.}~\bibnamefont
  {Liu}}, \bibinfo {author} {\bibfnamefont {Z.}~\bibnamefont {Wang}}, \bibinfo
  {author} {\bibfnamefont {K.}~\bibnamefont {Watanabe}}, \bibinfo {author}
  {\bibfnamefont {T.}~\bibnamefont {Taniguchi}}, \bibinfo {author}
  {\bibfnamefont {O.}~\bibnamefont {Vafek}}, \ and\ \bibinfo {author}
  {\bibfnamefont {J.}~\bibnamefont {Li}},\ }\href@noop {} {\bibfield  {journal}
  {\bibinfo  {journal} {Science}\ }\textbf {\bibinfo {volume} {371}},\ \bibinfo
  {pages} {1261} (\bibinfo {year} {2021})}\BibitemShut {NoStop}%
\bibitem [{\citenamefont {Cheng}\ \emph {et~al.}(2015)\citenamefont {Cheng},
  \citenamefont {Tomczyk}, \citenamefont {Lu}, \citenamefont {Veazey},
  \citenamefont {Huang}, \citenamefont {Irvin}, \citenamefont {Ryu},
  \citenamefont {Lee}, \citenamefont {Eom}, \citenamefont {Hellberg} \emph
  {et~al.}}]{cheng2015electron}%
  \BibitemOpen
  \bibfield  {author} {\bibinfo {author} {\bibfnamefont {G.}~\bibnamefont
  {Cheng}}, \bibinfo {author} {\bibfnamefont {M.}~\bibnamefont {Tomczyk}},
  \bibinfo {author} {\bibfnamefont {S.}~\bibnamefont {Lu}}, \bibinfo {author}
  {\bibfnamefont {J.~P.}\ \bibnamefont {Veazey}}, \bibinfo {author}
  {\bibfnamefont {M.}~\bibnamefont {Huang}}, \bibinfo {author} {\bibfnamefont
  {P.}~\bibnamefont {Irvin}}, \bibinfo {author} {\bibfnamefont
  {S.}~\bibnamefont {Ryu}}, \bibinfo {author} {\bibfnamefont {H.}~\bibnamefont
  {Lee}}, \bibinfo {author} {\bibfnamefont {C.-B.}\ \bibnamefont {Eom}},
  \bibinfo {author} {\bibfnamefont {C.~S.}\ \bibnamefont {Hellberg}},  \emph
  {et~al.},\ }\href@noop {} {\bibfield  {journal} {\bibinfo  {journal}
  {Nature}\ }\textbf {\bibinfo {volume} {521}},\ \bibinfo {pages} {196}
  (\bibinfo {year} {2015})}\BibitemShut {NoStop}%
\bibitem [{\citenamefont {Inbar}\ \emph {et~al.}(2023)\citenamefont {Inbar},
  \citenamefont {Birkbeck}, \citenamefont {Xiao}, \citenamefont {Taniguchi},
  \citenamefont {Watanabe}, \citenamefont {Yan}, \citenamefont {Oreg},
  \citenamefont {Stern}, \citenamefont {Berg},\ and\ \citenamefont
  {Ilani}}]{inbar2023quantum}%
  \BibitemOpen
  \bibfield  {author} {\bibinfo {author} {\bibfnamefont {A.}~\bibnamefont
  {Inbar}}, \bibinfo {author} {\bibfnamefont {J.}~\bibnamefont {Birkbeck}},
  \bibinfo {author} {\bibfnamefont {J.}~\bibnamefont {Xiao}}, \bibinfo {author}
  {\bibfnamefont {T.}~\bibnamefont {Taniguchi}}, \bibinfo {author}
  {\bibfnamefont {K.}~\bibnamefont {Watanabe}}, \bibinfo {author}
  {\bibfnamefont {B.}~\bibnamefont {Yan}}, \bibinfo {author} {\bibfnamefont
  {Y.}~\bibnamefont {Oreg}}, \bibinfo {author} {\bibfnamefont {A.}~\bibnamefont
  {Stern}}, \bibinfo {author} {\bibfnamefont {E.}~\bibnamefont {Berg}}, \ and\
  \bibinfo {author} {\bibfnamefont {S.}~\bibnamefont {Ilani}},\ }\href@noop {}
  {\bibfield  {journal} {\bibinfo  {journal} {Nature}\ }\textbf {\bibinfo
  {volume} {614}},\ \bibinfo {pages} {682} (\bibinfo {year}
  {2023})}\BibitemShut {NoStop}%
\bibitem [{\citenamefont {Kozuka}\ \emph {et~al.}(2010)\citenamefont {Kozuka},
  \citenamefont {Kim}, \citenamefont {Ohta}, \citenamefont {Hikita},
  \citenamefont {Bell},\ and\ \citenamefont {Hwang}}]{10.1063/1.3524198}%
  \BibitemOpen
  \bibfield  {author} {\bibinfo {author} {\bibfnamefont {Y.}~\bibnamefont
  {Kozuka}}, \bibinfo {author} {\bibfnamefont {M.}~\bibnamefont {Kim}},
  \bibinfo {author} {\bibfnamefont {H.}~\bibnamefont {Ohta}}, \bibinfo {author}
  {\bibfnamefont {Y.}~\bibnamefont {Hikita}}, \bibinfo {author} {\bibfnamefont
  {C.}~\bibnamefont {Bell}}, \ and\ \bibinfo {author} {\bibfnamefont {H.~Y.}\
  \bibnamefont {Hwang}},\ }\href {\doibase 10.1063/1.3524198} {\bibfield
  {journal} {\bibinfo  {journal} {Applied Physics Letters}\ }\textbf {\bibinfo
  {volume} {97}},\ \bibinfo {pages} {222115} (\bibinfo {year} {2010})},\
  \Eprint
  {http://arxiv.org/abs/https://pubs.aip.org/aip/apl/article-pdf/doi/10.1063/1.3524198/13506045/222115\_1\_online.pdf}
  {https://pubs.aip.org/aip/apl/article-pdf/doi/10.1063/1.3524198/13506045/222115\_1\_online.pdf}
  \BibitemShut {NoStop}%
\bibitem [{\citenamefont {Eugenio}\ and\ \citenamefont
  {Vafek}(2023)}]{paul2023}%
  \BibitemOpen
  \bibfield  {author} {\bibinfo {author} {\bibfnamefont {P.~M.}\ \bibnamefont
  {Eugenio}}\ and\ \bibinfo {author} {\bibfnamefont {O.}~\bibnamefont
  {Vafek}},\ }\href {\doibase 10.21468/SciPostPhys.15.3.081} {\bibfield
  {journal} {\bibinfo  {journal} {SciPost Phys.}\ }\textbf {\bibinfo {volume}
  {15}},\ \bibinfo {pages} {081} (\bibinfo {year} {2023})}\BibitemShut
  {NoStop}%
\bibitem [{\citenamefont {Kuklov}(1990)}]{Kuklov1990}%
  \BibitemOpen
  \bibfield  {author} {\bibinfo {author} {\bibfnamefont {A.}~\bibnamefont
  {Kuklov}},\ }\href@noop {} {\bibfield  {journal} {\bibinfo  {journal}
  {Superconductivity [translation from Russian journal, available upon request
  from the Niels Bohr Library, AIP.]}\ }\textbf {\bibinfo {volume} {3}},\
  \bibinfo {pages} {S335} (\bibinfo {year} {1990})}\BibitemShut {NoStop}%
\bibitem [{\citenamefont {Zhang}\ \emph {et~al.}(2024)\citenamefont {Zhang},
  \citenamefont {Kuklov}, \citenamefont {Prokof'ev},\ and\ \citenamefont
  {Svistunov}}]{zhang2024superconductivity}%
  \BibitemOpen
  \bibfield  {author} {\bibinfo {author} {\bibfnamefont {Z.}~\bibnamefont
  {Zhang}}, \bibinfo {author} {\bibfnamefont {A.}~\bibnamefont {Kuklov}},
  \bibinfo {author} {\bibfnamefont {N.}~\bibnamefont {Prokof'ev}}, \ and\
  \bibinfo {author} {\bibfnamefont {B.}~\bibnamefont {Svistunov}},\ }\href@noop
  {} {\bibfield  {journal} {\bibinfo  {journal} {arXiv preprint
  arXiv:2408.03266}\ } (\bibinfo {year} {2024})}\BibitemShut {NoStop}%
\bibitem [{\citenamefont {Kennes}\ \emph {et~al.}(2017)\citenamefont {Kennes},
  \citenamefont {Wilner}, \citenamefont {Reichman},\ and\ \citenamefont
  {Millis}}]{kennes2017transient}%
  \BibitemOpen
  \bibfield  {author} {\bibinfo {author} {\bibfnamefont {D.~M.}\ \bibnamefont
  {Kennes}}, \bibinfo {author} {\bibfnamefont {E.~Y.}\ \bibnamefont {Wilner}},
  \bibinfo {author} {\bibfnamefont {D.~R.}\ \bibnamefont {Reichman}}, \ and\
  \bibinfo {author} {\bibfnamefont {A.~J.}\ \bibnamefont {Millis}},\
  }\href@noop {} {\bibfield  {journal} {\bibinfo  {journal} {Nature Physics}\
  }\textbf {\bibinfo {volume} {13}},\ \bibinfo {pages} {479} (\bibinfo {year}
  {2017})}\BibitemShut {NoStop}%
\bibitem [{\citenamefont {Sous}\ \emph {et~al.}(2021)\citenamefont {Sous},
  \citenamefont {Kloss}, \citenamefont {Kennes}, \citenamefont {Reichman},\
  and\ \citenamefont {Millis}}]{sous2021phonon}%
  \BibitemOpen
  \bibfield  {author} {\bibinfo {author} {\bibfnamefont {J.}~\bibnamefont
  {Sous}}, \bibinfo {author} {\bibfnamefont {B.}~\bibnamefont {Kloss}},
  \bibinfo {author} {\bibfnamefont {D.~M.}\ \bibnamefont {Kennes}}, \bibinfo
  {author} {\bibfnamefont {D.~R.}\ \bibnamefont {Reichman}}, \ and\ \bibinfo
  {author} {\bibfnamefont {A.~J.}\ \bibnamefont {Millis}},\ }\href@noop {}
  {\bibfield  {journal} {\bibinfo  {journal} {Nature communications}\ }\textbf
  {\bibinfo {volume} {12}},\ \bibinfo {pages} {5803} (\bibinfo {year}
  {2021})}\BibitemShut {NoStop}%
\bibitem [{\citenamefont {He}\ and\ \citenamefont
  {Vanderbilt}(2001)}]{vanderbilt2001}%
  \BibitemOpen
  \bibfield  {author} {\bibinfo {author} {\bibfnamefont {L.}~\bibnamefont
  {He}}\ and\ \bibinfo {author} {\bibfnamefont {D.}~\bibnamefont
  {Vanderbilt}},\ }\href {\doibase 10.1103/PhysRevLett.86.5341} {\bibfield
  {journal} {\bibinfo  {journal} {Phys. Rev. Lett.}\ }\textbf {\bibinfo
  {volume} {86}},\ \bibinfo {pages} {5341} (\bibinfo {year}
  {2001})}\BibitemShut {NoStop}%
\bibitem [{\citenamefont {Marzari}\ \emph {et~al.}(2012)\citenamefont
  {Marzari}, \citenamefont {Mostofi}, \citenamefont {Yates}, \citenamefont
  {Souza},\ and\ \citenamefont {Vanderbilt}}]{wannierrev}%
  \BibitemOpen
  \bibfield  {author} {\bibinfo {author} {\bibfnamefont {N.}~\bibnamefont
  {Marzari}}, \bibinfo {author} {\bibfnamefont {A.~A.}\ \bibnamefont
  {Mostofi}}, \bibinfo {author} {\bibfnamefont {J.~R.}\ \bibnamefont {Yates}},
  \bibinfo {author} {\bibfnamefont {I.}~\bibnamefont {Souza}}, \ and\ \bibinfo
  {author} {\bibfnamefont {D.}~\bibnamefont {Vanderbilt}},\ }\href {\doibase
  10.1103/RevModPhys.84.1419} {\bibfield  {journal} {\bibinfo  {journal} {Rev.
  Mod. Phys.}\ }\textbf {\bibinfo {volume} {84}},\ \bibinfo {pages} {1419}
  (\bibinfo {year} {2012})}\BibitemShut {NoStop}%
\bibitem [{\citenamefont {Proville}\ and\ \citenamefont
  {Aubry}(1999)}]{proville1999small}%
  \BibitemOpen
  \bibfield  {author} {\bibinfo {author} {\bibfnamefont {L.}~\bibnamefont
  {Proville}}\ and\ \bibinfo {author} {\bibfnamefont {S.}~\bibnamefont
  {Aubry}},\ }\href@noop {} {\bibfield  {journal} {\bibinfo  {journal} {The
  European Physical Journal B-Condensed Matter and Complex Systems}\ }\textbf
  {\bibinfo {volume} {11}},\ \bibinfo {pages} {41} (\bibinfo {year}
  {1999})}\BibitemShut {NoStop}%
\bibitem [{\citenamefont {Proville}\ and\ \citenamefont
  {Aubry}(2000)}]{proville2000small}%
  \BibitemOpen
  \bibfield  {author} {\bibinfo {author} {\bibfnamefont {L.}~\bibnamefont
  {Proville}}\ and\ \bibinfo {author} {\bibfnamefont {S.}~\bibnamefont
  {Aubry}},\ }\href@noop {} {\bibfield  {journal} {\bibinfo  {journal} {The
  European Physical Journal B-Condensed Matter and Complex Systems}\ }\textbf
  {\bibinfo {volume} {15}},\ \bibinfo {pages} {405} (\bibinfo {year}
  {2000})}\BibitemShut {NoStop}%
\bibitem [{\citenamefont {Proville}\ and\ \citenamefont
  {Aubry}(1998)}]{PROVILLE1998307}%
  \BibitemOpen
  \bibfield  {author} {\bibinfo {author} {\bibfnamefont {L.}~\bibnamefont
  {Proville}}\ and\ \bibinfo {author} {\bibfnamefont {S.}~\bibnamefont
  {Aubry}},\ }\href {\doibase https://doi.org/10.1016/S0167-2789(97)00283-2}
  {\bibfield  {journal} {\bibinfo  {journal} {Physica D: Nonlinear Phenomena}\
  }\textbf {\bibinfo {volume} {113}},\ \bibinfo {pages} {307 } (\bibinfo {year}
  {1998})},\ \bibinfo {note} {proceedings of the Conference on Fluctuations,
  Nonlinearity and Disorder in Condensed Matter and Biological
  Physics}\BibitemShut {NoStop}%
\bibitem [{\citenamefont {Bari{\v{s}}i{\'c}}\ and\ \citenamefont
  {Bari{\v{s}}i{\'c}}(2012)}]{barivsic2012bipolarons}%
  \BibitemOpen
  \bibfield  {author} {\bibinfo {author} {\bibfnamefont {O.}~\bibnamefont
  {Bari{\v{s}}i{\'c}}}\ and\ \bibinfo {author} {\bibfnamefont {S.}~\bibnamefont
  {Bari{\v{s}}i{\'c}}},\ }\href@noop {} {\bibfield  {journal} {\bibinfo
  {journal} {The European Physical Journal B}\ }\textbf {\bibinfo {volume}
  {85}},\ \bibinfo {pages} {1} (\bibinfo {year} {2012})}\BibitemShut {NoStop}%
\bibitem [{\citenamefont {Bon\ifmmode~\check{c}\else \v{c}\fi{}a}\ and\
  \citenamefont {Trugman}(2001)}]{PhysRevB.64.094507}%
  \BibitemOpen
  \bibfield  {author} {\bibinfo {author} {\bibfnamefont {J.}~\bibnamefont
  {Bon\ifmmode~\check{c}\else \v{c}\fi{}a}}\ and\ \bibinfo {author}
  {\bibfnamefont {S.~A.}\ \bibnamefont {Trugman}},\ }\href {\doibase
  10.1103/PhysRevB.64.094507} {\bibfield  {journal} {\bibinfo  {journal} {Phys.
  Rev. B}\ }\textbf {\bibinfo {volume} {64}},\ \bibinfo {pages} {094507}
  (\bibinfo {year} {2001})}\BibitemShut {NoStop}%
\bibitem [{\citenamefont {Hague}\ and\ \citenamefont
  {Kornilovitch}(2009)}]{PhysRevB.80.054301}%
  \BibitemOpen
  \bibfield  {author} {\bibinfo {author} {\bibfnamefont {J.~P.}\ \bibnamefont
  {Hague}}\ and\ \bibinfo {author} {\bibfnamefont {P.~E.}\ \bibnamefont
  {Kornilovitch}},\ }\href {\doibase 10.1103/PhysRevB.80.054301} {\bibfield
  {journal} {\bibinfo  {journal} {Phys. Rev. B}\ }\textbf {\bibinfo {volume}
  {80}},\ \bibinfo {pages} {054301} (\bibinfo {year} {2009})}\BibitemShut
  {NoStop}%
\bibitem [{\citenamefont {Alexandrov}\ \emph {et~al.}(2012)\citenamefont
  {Alexandrov}, \citenamefont {Samson},\ and\ \citenamefont
  {Sica}}]{PhysRevB.85.104520}%
  \BibitemOpen
  \bibfield  {author} {\bibinfo {author} {\bibfnamefont {A.~S.}\ \bibnamefont
  {Alexandrov}}, \bibinfo {author} {\bibfnamefont {J.~H.}\ \bibnamefont
  {Samson}}, \ and\ \bibinfo {author} {\bibfnamefont {G.}~\bibnamefont
  {Sica}},\ }\href {\doibase 10.1103/PhysRevB.85.104520} {\bibfield  {journal}
  {\bibinfo  {journal} {Phys. Rev. B}\ }\textbf {\bibinfo {volume} {85}},\
  \bibinfo {pages} {104520} (\bibinfo {year} {2012})}\BibitemShut {NoStop}%
\bibitem [{\citenamefont {Hague}\ \emph
  {et~al.}(2007{\natexlab{a}})\citenamefont {Hague}, \citenamefont
  {Kornilovitch}, \citenamefont {Samson},\ and\ \citenamefont
  {Alexandrov}}]{hague2007superlight}%
  \BibitemOpen
  \bibfield  {author} {\bibinfo {author} {\bibfnamefont {J.}~\bibnamefont
  {Hague}}, \bibinfo {author} {\bibfnamefont {P.}~\bibnamefont {Kornilovitch}},
  \bibinfo {author} {\bibfnamefont {J.}~\bibnamefont {Samson}}, \ and\ \bibinfo
  {author} {\bibfnamefont {A.}~\bibnamefont {Alexandrov}},\ }\href@noop {}
  {\bibfield  {journal} {\bibinfo  {journal} {Journal of Physics: Condensed
  Matter}\ }\textbf {\bibinfo {volume} {19}},\ \bibinfo {pages} {255214}
  (\bibinfo {year} {2007}{\natexlab{a}})}\BibitemShut {NoStop}%
\bibitem [{\citenamefont {Hague}\ \emph
  {et~al.}(2007{\natexlab{b}})\citenamefont {Hague}, \citenamefont
  {Kornilovitch}, \citenamefont {Samson},\ and\ \citenamefont
  {Alexandrov}}]{PhysRevLett.98.037002}%
  \BibitemOpen
  \bibfield  {author} {\bibinfo {author} {\bibfnamefont {J.~P.}\ \bibnamefont
  {Hague}}, \bibinfo {author} {\bibfnamefont {P.~E.}\ \bibnamefont
  {Kornilovitch}}, \bibinfo {author} {\bibfnamefont {J.~H.}\ \bibnamefont
  {Samson}}, \ and\ \bibinfo {author} {\bibfnamefont {A.~S.}\ \bibnamefont
  {Alexandrov}},\ }\href {\doibase 10.1103/PhysRevLett.98.037002} {\bibfield
  {journal} {\bibinfo  {journal} {Phys. Rev. Lett.}\ }\textbf {\bibinfo
  {volume} {98}},\ \bibinfo {pages} {037002} (\bibinfo {year}
  {2007}{\natexlab{b}})}\BibitemShut {NoStop}%
\end{thebibliography}%


\begin{thebibliography}{4}%
\makeatletter
\providecommand \@ifxundefined [1]{%
 \@ifx{#1\undefined}
}%
\providecommand \@ifnum [1]{%
 \ifnum #1\expandafter \@firstoftwo
 \else \expandafter \@secondoftwo
 \fi
}%
\providecommand \@ifx [1]{%
 \ifx #1\expandafter \@firstoftwo
 \else \expandafter \@secondoftwo
 \fi
}%
\providecommand \natexlab [1]{#1}%
\providecommand \enquote  [1]{``#1''}%
\providecommand \bibnamefont  [1]{#1}%
\providecommand \bibfnamefont [1]{#1}%
\providecommand \citenamefont [1]{#1}%
\providecommand \href@noop [0]{\@secondoftwo}%
\providecommand \href [0]{\begingroup \@sanitize@url \@href}%
\providecommand \@href[1]{\@@startlink{#1}\@@href}%
\providecommand \@@href[1]{\endgroup#1\@@endlink}%
\providecommand \@sanitize@url [0]{\catcode `\\12\catcode `\$12\catcode
  `\&12\catcode `\#12\catcode `\^12\catcode `\_12\catcode `\%12\relax}%
\providecommand \@@startlink[1]{}%
\providecommand \@@endlink[0]{}%
\providecommand \url  [0]{\begingroup\@sanitize@url \@url }%
\providecommand \@url [1]{\endgroup\@href {#1}{\urlprefix }}%
\providecommand \urlprefix  [0]{URL }%
\providecommand \Eprint [0]{\href }%
\providecommand \doibase [0]{http://dx.doi.org/}%
\providecommand \selectlanguage [0]{\@gobble}%
\providecommand \bibinfo  [0]{\@secondoftwo}%
\providecommand \bibfield  [0]{\@secondoftwo}%
\providecommand \translation [1]{[#1]}%
\providecommand \BibitemOpen [0]{}%
\providecommand \bibitemStop [0]{}%
\providecommand \bibitemNoStop [0]{.\EOS\space}%
\providecommand \EOS [0]{\spacefactor3000\relax}%
\providecommand \BibitemShut  [1]{\csname bibitem#1\endcsname}%
\let\auto@bib@innerbib\@empty
\bibitem [{\citenamefont {He}\ and\ \citenamefont
  {Vanderbilt}(2001)}]{vanderbilt2001}%
  \BibitemOpen
  \bibfield  {author} {\bibinfo {author} {\bibfnamefont {L.}~\bibnamefont
  {He}}\ and\ \bibinfo {author} {\bibfnamefont {D.}~\bibnamefont
  {Vanderbilt}},\ }\href {\doibase 10.1103/PhysRevLett.86.5341} {\bibfield
  {journal} {\bibinfo  {journal} {Phys. Rev. Lett.}\ }\textbf {\bibinfo
  {volume} {86}},\ \bibinfo {pages} {5341} (\bibinfo {year}
  {2001})}\BibitemShut {NoStop}%
\bibitem [{\citenamefont {Marzari}\ \emph {et~al.}(2012)\citenamefont
  {Marzari}, \citenamefont {Mostofi}, \citenamefont {Yates}, \citenamefont
  {Souza},\ and\ \citenamefont {Vanderbilt}}]{wannierrev}%
  \BibitemOpen
  \bibfield  {author} {\bibinfo {author} {\bibfnamefont {N.}~\bibnamefont
  {Marzari}}, \bibinfo {author} {\bibfnamefont {A.~A.}\ \bibnamefont
  {Mostofi}}, \bibinfo {author} {\bibfnamefont {J.~R.}\ \bibnamefont {Yates}},
  \bibinfo {author} {\bibfnamefont {I.}~\bibnamefont {Souza}}, \ and\ \bibinfo
  {author} {\bibfnamefont {D.}~\bibnamefont {Vanderbilt}},\ }\href {\doibase
  10.1103/RevModPhys.84.1419} {\bibfield  {journal} {\bibinfo  {journal} {Rev.
  Mod. Phys.}\ }\textbf {\bibinfo {volume} {84}},\ \bibinfo {pages} {1419}
  (\bibinfo {year} {2012})}\BibitemShut {NoStop}%
\bibitem [{\citenamefont {Volkov}\ \emph {et~al.}(2022)\citenamefont {Volkov},
  \citenamefont {Chandra},\ and\ \citenamefont
  {Coleman}}]{volkov2022superconductivity}%
  \BibitemOpen
  \bibfield  {author} {\bibinfo {author} {\bibfnamefont {P.~A.}\ \bibnamefont
  {Volkov}}, \bibinfo {author} {\bibfnamefont {P.}~\bibnamefont {Chandra}}, \
  and\ \bibinfo {author} {\bibfnamefont {P.}~\bibnamefont {Coleman}},\ }\href
  {\doibase 10.1038/s41467-022-32303-2} {\bibfield  {journal} {\bibinfo
  {journal} {Nature communications}\ }\textbf {\bibinfo {volume} {13}},\
  \bibinfo {pages} {4599} (\bibinfo {year} {2022})}\BibitemShut {NoStop}%
\bibitem [{\citenamefont {B{\"a}uerle}\ \emph {et~al.}(1980)\citenamefont
  {B{\"a}uerle}, \citenamefont {Wagner}, \citenamefont {W{\"o}hlecke},
  \citenamefont {Dorner},\ and\ \citenamefont
  {Kraxenberger}}]{bauerle1980soft}%
  \BibitemOpen
  \bibfield  {author} {\bibinfo {author} {\bibfnamefont {D.}~\bibnamefont
  {B{\"a}uerle}}, \bibinfo {author} {\bibfnamefont {D.}~\bibnamefont {Wagner}},
  \bibinfo {author} {\bibfnamefont {M.}~\bibnamefont {W{\"o}hlecke}}, \bibinfo
  {author} {\bibfnamefont {B.}~\bibnamefont {Dorner}}, \ and\ \bibinfo {author}
  {\bibfnamefont {H.}~\bibnamefont {Kraxenberger}},\ }\href@noop {} {\bibfield
  {journal} {\bibinfo  {journal} {Zeitschrift f{\"u}r Physik B Condensed
  Matter}\ }\textbf {\bibinfo {volume} {38}},\ \bibinfo {pages} {335} (\bibinfo
  {year} {1980})}\BibitemShut {NoStop}%
\end{thebibliography}%

\end{document}


\newcommand{\pv}[1]{{\bf \color{red}PV: }{\color{blue} #1}}

\newcommand{\ZH}[1]{{\bf \color{red}ZH: }{\color{red} #1}}

\title{``Quantum bipolaron'' superconductivity from quadratic electron-phonon coupling: Supplemental Materials}
\author{Zhaoyu Han}
\affiliation{Department of Physics, Stanford University, Stanford, California 94305, USA}

\author{Steven A. Kivelson}
\affiliation{Department of Physics, Stanford University, Stanford, California 94305, USA}

\author{Pavel A. Volkov}

\affiliation{Department of Physics, University of Connecticut, Storrs, Connecticut 06269, USA}
\affiliation{Department of Physics, Harvard University, Cambridge, Massachusetts 02138, USA}
\maketitle

\section{Derivation of the effective coefficients}

In this work, we study the simplest model that exhibits non-linear e-ph coupling, the quadratic Holstein model:
\begin{align}\label{eq: model}
    \hat{H} = & -t \sum_{\langle ij \rangle \sigma } \left(\hat{c}^\dagger_{i\sigma}\hat{c}_{j\sigma} + \text{h.c.}\right) \nonumber\\
    & \ \ + \sum_{i} \left[\frac{K}{2} (1+g \hat{n}_i)\hat{X}_i^2 +\frac{\hat{P}_i^2}{2M}\right]
\end{align}
where we have assumed that the electron density on each site is coupled to the square of the phonon coordinate on that site. The dimensionless factor $g$ thus quantifies the degree of e-ph coupling, and $g>-1/2$ for the validity of this model (for the case of $g<-1/2$, higher order terms in phonon potential should be considered). On a site with $m=0,1,2$ electrons on it, the phonon oscillates with frequency $\omega_m \equiv \sqrt{1+mg} \omega_0$ where $\omega_0 \equiv \sqrt{K/M}$ is the characteristic frequency of the system. 

We find that, just like its linear counterpart, the quadratic e-ph coupling still mediates an effective electron-electron attraction, albeit due to a completely different mechanism. To see this, we may neglect $t$ and consider the two-electron sector of the problem: there are two ways to arrange the two electrons, one is to assign them to the same site, and the other is to put them on two different sites. As opposed to the linear case, the two types of states have the same energy at the classical ($M\rightarrow \infty$) limit. However, they are distinguished by the difference in zero-point energy for any finite ion mass $M$. Specifically, the energy of a doubly occupied site and an empty site  is lower than that of two singly occupied sites by an amount
\begin{align}
    \Ueph \equiv& \frac{1}{2}\left(2\omega_1-\omega_2-\omega_0\right) \\
    =& \omega_0\left(\sqrt{1+g}-\frac{\sqrt{1+2g}+1}{2}\right).
\end{align}
Importantly, $\Ueph>0$ for any $g>-1/2$ (since $\omega_1 = \sqrt{(\omega_0^2+\omega_2^2)/2}>(\omega_0+\omega_2)/2$), which makes the formation of bipolaron always favorable.

For a site with $m=0,1,2$ electrons on it, we denote the $n$-th phonon eigen-wavefunctions as $\phi^{m}_n(x)$, whose expression is easy to obtain:
\begin{align}
    \phi^{(m)}_n(x) = \frac{1}{\sqrt{2^n n! W_m}} \mathrm{e}^{-x^2/(2W_m^2)} H_n(x/W_m)
\end{align}
where $W_m \equiv 1/\sqrt[4]{MK(1+mg)}$ is the width of the wavepacket, and $H_n$ is the $n$-th Hermite polynomial.

With perturbation theory treating $t$ as a small parameter, we are able to obtain the expression for the bipolaron hopping $\tau$ as defined in the main text. To do so, we may consider two neighboring sites, and hop the bipolaron through all possible virtual processes, where all the intermediate states with $2n_1$ and $2n_2$ phonons on the two sites need to be considered ($n_1,n_2$ are non-negative integers). We note that since we start from phonon ground state with even parity, we only need to consider intermediate states with an even number of phonons on each site. Thus, we get the expression:
\begin{align}
    \tau =& \sum_{n_1 n_2}\frac{2t^2}{\Ueph+2(n_1+n_2)\omega_1} \langle \phi^{(2)}_0|\phi^{(1)}_{2n_1}\rangle  \langle \phi^{(1)}_{2n_1}|\phi^{(0)}_0\rangle  \cdot  \langle \phi^{(0)}_0|\phi^{(1)}_{2n_2}\rangle  \langle \phi^{(1)}_{2n_2}|\phi^{(2)}_0\rangle \\
    =&\sum_{n_1 n_2}\frac{2t^2}{\Ueph+2(n_1+n_2)\omega_1} \frac{\sqrt{2W_0W_1}}{\sqrt{W_0^2+W_1^2}} \frac{(2n_1)!}{4^{n_1} (n_1!)^2}(\gamma_0\gamma_2)^{n_1}\frac{\sqrt{2W_2W_1}}{\sqrt{W_2^2+W_1^2}}  \cdot \frac{\sqrt{2W_2W_1}}{\sqrt{W_2^2+W_1^2}} \frac{(2n_2)!}{4^{n_2} (n_2!)^2}(\gamma_0\gamma_2)^{n_2} \frac{\sqrt{2W_0W_1}}{\sqrt{W_0^2+W_1^2}}  \\
    =&\sum_{n_1 n_2}\frac{2\eta t^2}{\Ueph+2(n_1+n_2)\omega_1} \frac{(2n_1)!}{4^{n_1} (n_1!)^2}(\gamma_0\gamma_2)^{n_1}  \cdot  \frac{(2n_2)!}{4^{n_2} (n_2!)^2}(\gamma_0\gamma_2)^{n_2} 
\end{align}
where
\begin{align}
    \gamma_{m} & = \frac{\sqrt{1+g}-\sqrt{1+mg}}{\sqrt{1+g}+\sqrt{1+mg}} \\
    \eta &\equiv  \frac{2W_0W_1}{W_0^2+W_1^2} \frac{2W_2W_1}{W_2^2+W_1^2} = \frac{4\sqrt[4]{1+2g}\sqrt{1+g} }{\left(1+\sqrt{1+g}\right)\left(\sqrt{1+2 g}+\sqrt{1+g}\right)}\\
    \xi &\equiv \omega_1/\Ueph = \frac{2\sqrt{1+g}}{2\sqrt{1+g}-\sqrt{1+2g}-1}
\end{align}
Then we use Feynman's trick $1/A = \int_0^\infty dx \mathrm{e}^{-A x}$ and continue derivation:
\begin{align}
    \tau = &\frac{2\eta t^2}{\Ueph} \sum_{n_1 n_2}\int_0^\infty dx \mathrm{e}^{-x} \frac{(2n_1)!}{4^{n_1} (n_1!)^2}(\gamma_0\gamma_2 \mathrm{e}^{-2\xi x})^{n_1}  \cdot  \frac{(2n_2)!}{4^{n_2} (n_2!)^2}(\gamma_0\gamma_2 \mathrm{e}^{-2\xi x})^{n_2} \\
    =& \frac{2\eta t^2}{\Ueph} \int_0^\infty dx \mathrm{e}^{-x} \frac{1}{\sqrt{1-\gamma_0\gamma_2 \mathrm{e}^{-2\xi x}}}  \cdot  \frac{1}{\sqrt{1-\gamma_0\gamma_2 \mathrm{e}^{-2\xi x}}} \\
    =& \frac{2\eta t^2}{\Ueph} \int_0^\infty dx \mathrm{e}^{-x} \frac{1}{1-\gamma_0\gamma_2 \mathrm{e}^{-2\xi x}} 
\end{align}

Similarly, we are able to derive the expression for $V$:
\begin{align}
     V =& \sum_{n_1 n_2}\frac{4t^2}{\Ueph+2(n_1+n_2)\omega_1} \langle \phi^{(2)}_0|\phi^{(1)}_{2n_1}\rangle  \langle \phi^{(1)}_{2n_1}|\phi^{(2)}_0\rangle  \cdot  \langle \phi^{(0)}_0|\phi^{(1)}_{2n_2}\rangle  \langle \phi^{(1)}_{2n_2}|\phi^{(0)}_0\rangle \\
    =&\sum_{n_1 n_2}\frac{4t^2}{\Ueph+2(n_1+n_2)\omega_1} \frac{\sqrt{2W_2W_1}}{\sqrt{W_2^2+W_1^2}} \frac{(2n_1)!}{4^{n_1} (n_1!)^2}(\gamma_2^2)^{n_1}\frac{\sqrt{2W_2W_1}}{\sqrt{W_2^2+W_1^2}}  \cdot \frac{\sqrt{2W_0W_1}}{\sqrt{W_0^2+W_1^2}} \frac{(2n_2)!}{4^{n_2} (n_2!)^2}(\gamma_0^2)^{n_2} \frac{\sqrt{2W_0W_1}}{\sqrt{W_0^2+W_1^2}}  \\
    =&\sum_{n_1 n_2}\frac{4\eta t^2}{\Ueph+2(n_1+n_2)\omega_1} \frac{(2n_1)!}{4^{n_1} (n_1!)^2}(\gamma_2^2)^{n_1}  \cdot  \frac{(2n_2)!}{4^{n_2} (n_2!)^2}(\gamma_0^2)^{n_2} \\
    =&\frac{4\eta t^2}{\Ueph} \sum_{n_1 n_2}\int_0^\infty dx \mathrm{e}^{-x} \frac{(2n_1)!}{4^{n_1} (n_1!)^2}(\gamma_2^2 \mathrm{e}^{-2\xi x})^{n_1}  \cdot  \frac{(2n_2)!}{4^{n_2} (n_2!)^2}(\gamma_0^2 \mathrm{e}^{-2\xi x})^{n_2} \\
    =& \frac{4\eta t^2}{\Ueph} \int_0^\infty dx \mathrm{e}^{-x} \frac{1}{\sqrt{1-\gamma_2^2 \mathrm{e}^{-2\xi x}}}  \cdot  \frac{1}{\sqrt{1-\gamma_0^2 \mathrm{e}^{-2\xi x}}}
\end{align}

In the main text, for convenience, we introduced the definition of dimensionless parameters:
\begin{align}
    F_\tau(g) \equiv \tau/(t^2/\Ueph)\ , \ \ F_V(g) \equiv V/(t^2/\Ueph)
\end{align}


\subsection{ $g\rightarrow \infty$ limit}

In this limit, we have
\begin{align}
   &\xi \rightarrow 1/(1-\sqrt{2}/2) \approx 3.414 \\
   &\gamma_0 \rightarrow 1 \ , \ \ \gamma_2 \rightarrow 2\sqrt{2}-3\approx -0.172 \\
   & \eta \rightarrow \frac{4\sqrt[4]{2}}{1+\sqrt{2}} g^{-1/4}\approx 1.97 g^{-1/4}\\
   & \int_0^\infty dx \mathrm{e}^{-x} \frac{1}{1-\gamma_0\gamma_2 \mathrm{e}^{-2\xi x}}  \rightarrow 0.979 \\
   & \int_0^\infty dx \mathrm{e}^{-x} \frac{1}{\sqrt{1-\gamma_2^2 \mathrm{e}^{-2\xi x}}}  \cdot  \frac{1}{\sqrt{1-\gamma_0^2 \mathrm{e}^{-2\xi x}}} \rightarrow 1.194
\end{align}
Therefore, $\tau \rightarrow 3.861 g^{-1/4} \frac{t^2}{\omega_0}$ and $V \rightarrow 9.410 g^{-1/4} \frac{t^2}{\Ueph}$ in the $g\rightarrow \infty$ limit.

\subsection{ $g\rightarrow -1/2$ limit}

In this limit, we have
\begin{align}
   &\xi \rightarrow 1/(1-\sqrt{2}/2) \approx 3.414 \\
   &\gamma_0 \rightarrow 2\sqrt{2}-3\approx -0.172 \ , \ \ \gamma_2 \rightarrow 1 \\
   & \eta \rightarrow \frac{8 \sqrt[4]{2}}{2+\sqrt{2}} (g+1/2)^{1/4}\approx 2.79(g+1/2)^{1/4}\\
   & \int_0^\infty dx \mathrm{e}^{-x} \frac{1}{1-\gamma_0\gamma_2 \mathrm{e}^{-2\xi x}}  \rightarrow 0.979 \\
   & \int_0^\infty dx \mathrm{e}^{-x} \frac{1}{\sqrt{1-\gamma_2^2 \mathrm{e}^{-2\xi x}}}  \cdot  \frac{1}{\sqrt{1-\gamma_0^2 \mathrm{e}^{-2\xi x}}} \rightarrow 1.194
\end{align}
Therefore, $\tau \rightarrow 5.46 (g+1/2)^{1/4} \frac{t^2}{\Ueph}$ and $V \rightarrow 13.31 (g+1/2)^{1/4} \frac{t^2}{\Ueph}$ in the $g\rightarrow -1/2$ limit.



\section{Nonlinear e-ph coupling in a superlattice}

Let us first make a few general comments about electronic structure in the presence of a superlattice. Assuming the superlattice period $a_\text{sl}$ is much larger than the microscopic lattice one, $a_0$, one can describe the bands near top or bottom of the band with a continuum approximation:
\begin{equation}
\hat{H} = -\frac{\hbar^2 \Delta}{2m^*}+V_0 f_{sl}({\bf r}),
\end{equation}
where $|f_{sl}({\bf r})|<1$ describes the superlattice potential profile and $V_0$ - its magnitude. For concreteness, we focus on the case $m^*>0$, corresponding to the band bottom. Then, for $V_0\gg \frac{\hbar^2}{m^* a_\text{sl}^2}$ one expects the electronic states in the lowest superlattice band to form tightly localized Wannier states around the minima of $ f_{sl}({\bf r})$. An estimate of the orbital size can be obtained by expanding $ f_{sl}({\bf r})$ near the minimum, resulting in a harmonic oscillator problem. The ground state wavefunction's size is $a_\text{orb} \sim \sqrt{a_\text{sl}\frac{\hbar}{m^* V_0}}$. On the other hand, the hopping should generically scale exponentially with distance \cite{vanderbilt2001,wannierrev} leading to $t_{sl}\sim \frac{ \hbar^2}{m^* a_\text{sl}^2} e^{-a_\text{sl}/a_\text{orb}}$, which is suppressed exponentially strongly $\propto e^{-C \sqrt{V_0/(\hbar^2/m*a_\text{sl}^2)}}$ for $V_0\gg \frac{\hbar^2}{m*a_\text{sl}^2}$.

To consider the effects of quadratic electron-phonon coupling in such a setup, we consider a generalization of the model in Eq.~\ref{eq: model}:
\begin{align}\label{eq: N phonon model}
    \hat{H} = & -t \sum_{\langle ij \rangle \sigma } \left(\hat{c}^\dagger_{i\sigma}\hat{c}_{j\sigma} + \text{h.c.}\right) \nonumber\\
    & \ \ + \sum_{i}\sum_{\alpha=1}^{N} \left[\frac{K}{2} (1+g \hat{n}_i/N)\hat{X}_{i,\alpha}^2 +\frac{\hat{P}_{i,\alpha}^2}{2M}\right]
\end{align}
where we assumed that each Wannier orbital is uniformly coupled to $N$ microscopic phonon modes.

\begin{figure}[t]
    \centering
    \includegraphics[width=0.4\linewidth]{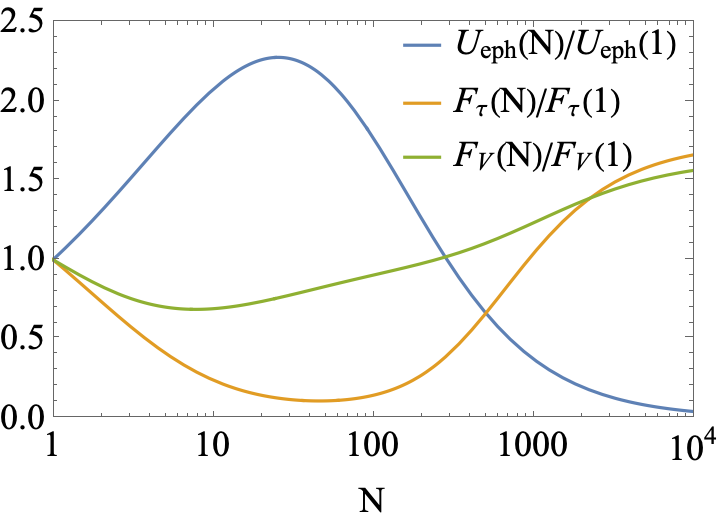}
    \caption{For the model where each Wannier orbital is uniformly coupled to $N$ phonons (Eq.~\ref{eq: N phonon model}), we plot the ratios $\Ueph(N)/\Ueph(1)$, $F_\tau(N)/F_\tau(1)$, and $F_V(N)/F_V(1)$, as a function of $N$ for a fixed $g=87.5$. In can be seen from the figure that, for this value of $g$, as long as $N$ is not as large as $10^4$, all the ratios are $\mathcal{O}(1)$. }
    \label{fig: superlattice}
\end{figure}

In the atomic limit $t=0$, the energy of one doubly occupied site and one empty site is lower than that of two singly occupied sites by an amount (binding energy): 
\begin{equation}
    \Ueph(N) = N \hbar\omega_0 \left(\sqrt{1+\frac{g}{N}} - \frac{\sqrt{1+\frac{2g}{N}}+1}{2}\right).
\end{equation}
Apparently, $ \Ueph(N=1)=\Ueph$ above, and its dependence on $N$ is plotted in Fig.~\ref{fig: superlattice}.

While for $g/N\ll1$ we get the result $\Ueph(N) = \Ueph/N$, in the opposite limit $g/N\gg1$  $\Ueph(N) = \sqrt{N} \Ueph$, i.e. introducing superlattice strengthens the coupling. Actually, $\Ueph(N)$ has a maximum at an intermediate value of $N$. For $g=87.5$, the maximum is obtained at $N\approx 25$ and is equal to $2.28 \Ueph$. Moreover, while $\Ueph(N)$ decreases monotonically afterwards, it reaches $\Ueph$ only at $N\approx300$. Such an orbital size corresponds to a unit cell much larger than $17\times 17$, which would be $6\times 6$ nm in the case of SrTiO$_3$.

Next, we turn to the evaluation of the effective coefficients:
\begin{align}
    \tau =& \sum_{\{n_i\},\{m_i\}}\frac{2t^2}{\Ueph(N)+\sum_{\alpha =1}^N 2(n_{1,\alpha }+n_{2,\alpha})\omega_1(N)} \prod_{\alpha=1}^{N} \left[ \langle \phi^{(2)}_0|\phi^{(1)}_{2n_{1,\alpha}}\rangle  \langle \phi^{(1)}_{2n_{1,\alpha}}|\phi^{(0)}_0\rangle  \cdot  \langle \phi^{(0)}_0|\phi^{(1)}_{2n_{2,\alpha}}\rangle  \langle \phi^{(1)}_{2n_{2,\alpha}}|\phi^{(2)}_0\rangle \right] \\
    =& \frac{2[\eta(N)]^N t^2}{\Ueph(N)} \int_0^\infty dx \mathrm{e}^{-x} \left[\frac{1}{1-\gamma_0(N)\gamma_2(N) \mathrm{e}^{-2\xi(N) x}} \right]^N \\
     V =& \frac{2[\eta(N)]^N t^2}{\Ueph(N)}  \int_0^\infty dx \mathrm{e}^{-x} \left[\frac{1}{\sqrt{1-[\gamma_2(N)]^2 \mathrm{e}^{-2\xi(N) x}}}  \cdot  \frac{1}{\sqrt{1-[\gamma_0(N)]^2 \mathrm{e}^{-2\xi(N) x}}}\right]^N
\end{align}
where
\begin{align}
    \gamma_{m}(N) & = \frac{\sqrt{1+g/N}-\sqrt{1+mg/N}}{\sqrt{1+g/N}+\sqrt{1+mg/N}} \\
    \eta(N) &\equiv  \frac{4\sqrt[4]{1+2g/N}\sqrt{1+g/N} }{\left(1+\sqrt{1+g/N}\right)\left(\sqrt{1+2 g/N}+\sqrt{1+g/N}\right)}\\
    \xi(N) &\equiv \omega_1(N) /\Ueph(N) = \frac{2\sqrt{1+g/N}}{N(2\sqrt{1+g/N}-\sqrt{1+2g/N}-1)}
\end{align}

Similar to the above, we still define:
\begin{align}
    F_\tau(g,N) \equiv \tau/[t^2/\Ueph(N)]\ , \ \ F_V(g,N) \equiv V/[t^2/\Ueph(N)]
\end{align}
and plot their dependence on $N$ for $g=87.5$ in Fig.~\ref{fig: superlattice}.

\section{Estimate of $g$ for SrTiO$_3$}

Converting it to the dimensionless $g$ we get $g \approx 2 (g^*/ a_0^3)\Omega_0^2/\Omega_T^2 \approx 87.5$ (using notations of \cite{volkov2022superconductivity} and $g* = 0.56 a_0^3$ from \cite{bauerle1980soft}; note that due to considerable dispersion of the TO phonon near $q=0$ we use its zone-edge energy rather then the zone-center one). The corresponding $\Ueph$ value is then around 50 meV.

Note that this result does not contradict the weak value of the resulting BCS coupling constant. The latter is of the order $(g^*/a_0^3)^2 \frac{\Omega_0^4}{\Omega_T^3} \rho_0  = (g^*/a_0^3)^2 \frac{\Omega_0^4}{\Omega_T^3 t} (k_F a_0)$. With $\Omega_0\approx 200$ meV, $\Omega_T\approx 20$ meV, bandwidth of the order $1 eV$ and small densities $k_F a_0 \sim 0.01$ we obtain the BCS couplings smaller then 1 (see \cite{volkov2022superconductivity} for more detailed estimates).

\bibliographystyle{apsrev4-1} 
\bibliography{ref}